\documentclass[%
 reprint,
 superscriptaddress,
 amsmath,amssymb,
 aps,
]{revtex4-2}

\usepackage{amsmath,amsthm}
\usepackage{graphicx}
\usepackage{dcolumn} 
\usepackage{bm}
\usepackage{braket}
\usepackage{todonotes}
\usepackage[compat=0.4]{yquant}
\usepackage{textgreek}
\usepackage{enumerate}

\newcommand{\abs}{\mathrm{abs}}
\newcommand{\Nw}{N_{\mathrm{w}}}
\newcommand{\Nspawn}{N_{\mathrm{spawn}}}
\newcommand{\vertiii}[1]{{\left\vert\kern-0.25ex\left\vert\kern-0.25ex\left\vert #1 
    \right\vert\kern-0.25ex\right\vert\kern-0.25ex\right\vert}}

\usepackage{xcolor}
\definecolor{riverlane_green}{RGB}{0, 111, 98}
\definecolor{riverlane_light_green}{RGB}{0, 150, 143}
\definecolor{riverlane_orange}{RGB}{255, 117, 0}
\definecolor{riverlane_red}{RGB}{220, 68, 5}
\definecolor{riverlane_pink}{RGB}{207, 111, 127}
\usepackage{hyperref}
\hypersetup{
  colorlinks   = true, 
  urlcolor     = riverlane_green, 
  linkcolor    = riverlane_orange, 
  citecolor   = riverlane_green  
}

\usepackage{tikz}
\usetikzlibrary{quantikz}

\usepackage{bm}
\usepackage{bbm}
\usepackage{bbold}

\begin{document}

\preprint{APS/123-QED}

\title{A Monte Carlo approach to bound Trotter error}

\author{Nick S. Blunt}
\email{nick.blunt@riverlane.com}
\affiliation{Riverlane, Cambridge, CB2 3BZ, United Kingdom}
\author{Aleksei V. Ivanov}
\affiliation{Riverlane, Cambridge, CB2 3BZ, United Kingdom}
\author{Andreas Juul Bay-Smidt}
\affiliation{NNF Quantum Computing Programme, Niels Bohr Institute, University of Copenhagen, Denmark}

\date{\today}

\begin{abstract}
Trotter product formulas are a natural and powerful approach to perform quantum simulation. However, the error analysis of product formulas is challenging, and their cost is often overestimated. It is established that Trotter error can be bounded in terms of spectral norms of nested commutators of the Hamiltonian partitions [Childs \emph{et al.}, Phys. Rev. X {\bf 11}, 011020], but evaluating these expressions is challenging, often achieved by repeated application of the triangle inequality, significantly loosening the bound. Here, we show that the spectral norm of an operator can be upper bounded by the spectral norm of an equivalent sign-problem-free operator, which can be calculated efficiently to large system sizes using projector Monte Carlo simulation. For a range of Hamiltonians and considering second-order formulas, we demonstrate that this Monte Carlo-based bound is often extremely tight, and even exact in some instances. For the uniform electron gas we reduce the cost of performing Trotterization from the literature by an order of magnitude. For the Pariser–Parr–Pople model for linear acene molecules, which has $\mathcal{O}(N^2)$ long-range interaction terms, we show that it suffices to use $\mathcal{O}(N^{0.57})$ Trotter steps and circuit depth $\mathcal{O}(N^{1.57})$ to implement Hamiltonian simulation. We hope that this approach will lead to a better understanding of the potential accuracy of Trotterization in a range of important applications.
\end{abstract}

\maketitle

\section{Introduction}
\label{sec:Introduction}

Quantum simulation is among the most promising applications of quantum computers, and one of the original motivations of quantum computation \cite{feynman_1982}. Among several quantum simulation methods, Trotterization is a simple and natural approach. Trotter product formulas, known as ``splitting methods'' in mathematics, have a long history and applications in a variety of numerical methods, both in classical and quantum simulation \cite{trotter_1959, suzuki_1985, suzuki_1990, suzuki_1991, lloyd_1996, babbush_2015, reiher_2017, childs_2018, childs_2019, kivlichan_2020}.

Trotterization has a number of features that make it appealing as a practical quantum simulation method. The quantum circuits to implement product formulas are particularly simple, requiring no ancilla qubits, and usually consisting of a small number of repeating layers of single-qubit rotation gates and basic entangling operations. These circuits can have some degree of tolerance to errors \cite{granet_2025}. Even on current quantum devices with high error rates, Trotterization can be used to obtain results at the frontier of state-of-the-art classical methods \cite{quantinuum_trotter, kim_2023}. The required number of Trotter steps to reach a given precision can scale sublinearly with system size, outperforming post-Trotter methods for some Hamiltonians \cite{childs_2019, childs_2021, kan_2025, bay_smidt_2025}. The precision of product formulas can also be exponentially improved using multi-product formulas or extrapolation methods \cite{childs_2012, low_2019, zhuk_2024, watson_2025}.

In the fault-tolerant setting, resource estimation studies of phase estimation have often favored qubitization~\cite{low_hamiltonian_2019,berry_improved_2018,poulin_quantum_algorithm_2019} over Trotterization, as assessed by non-Clifford gate counts \cite{babbush_2018_2, blunt_2022}. However, there are reasons to believe that the performance of Trotterization is significantly underestimated. First, Trotter circuits can be efficiently parallelized, reducing circuit depths and volumes, which is not accounted for by considering only non-Clifford gate counts. This point is particularly relevant given recent work on improved magic state distillation protocols \cite{cultivation}, which should allow efficient implementation of parallel rotation gates in a fault-tolerant architecture \cite{hirano_2025}. Second, the analysis of Trotter error is very challenging, and the error of product formulas is often overestimated, in some cases by multiple orders of magnitude.

A significant breakthrough showed that the worst-case Trotter error can be bounded in general by the spectral norm of nested commutators of the Hamiltonian partitions \cite{childs_2021}. However, numerically evaluating such expressions remains challenging, as it requires calculating the largest singular value of operators with support on an exponentially large Hilbert space, in addition to potential difficulties in even constructing such nested commutators. As such, a common approach is to instead apply the triangle inequality to simplify this task. Depending on how this is performed, this can significantly loosen the bound, overestimating the cost of Trotterization.

\begin{figure*}[t]
\includegraphics[width=1.0\linewidth]{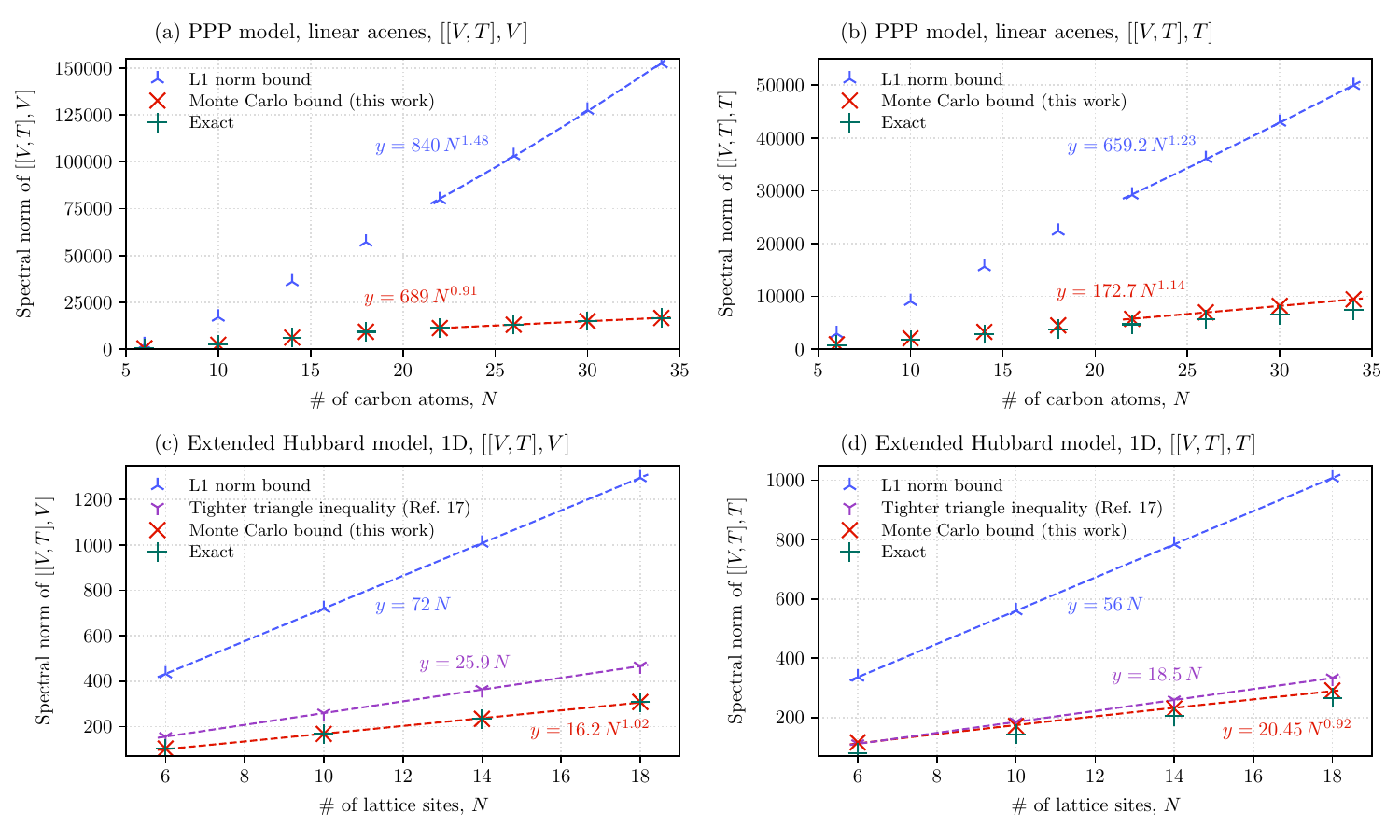}
\caption{Comparison of commutator bounds for PPP model and extended Hubbard model examples. In (a) and (b) we consider the PPP model applied to linear acene molecules, while (c) and (d) consider an extended Hubbard model on a one-dimensional lattice. The commutators considered are $[[V,T],V]$ and $[[V,T],T]$, for which we aim to provide tight upper bounds on the spectral norm (restricting to half-filling and projected spin $s_z=0$). Results in blue show an upper bound from the L1 norm of the commutator, Eq.~(\ref{eq:l1_norm}). Results in purple ((c) and (d) only) use a tighter triangle inequality from Ref.~\cite{bay_smidt_2025}. Results in red are calculated from the Monte Carlo bound. Exact results are obtained from either exact diagonalization or DMRG.}
\label{fig:1d}
\end{figure*}

This paper introduces a practical numerical approach to provide an upper bound for such Trotter error norms, using a novel strategy to avoid further application of the triangle inequality. We instead use that the spectral norm of any operator is upper bounded by the spectral norm of an equivalent sign-problem-free operator. Due to this sign-problem-free property, the latter norm can be evaluated efficiently by projector Monte Carlo methods. We show through a number of examples of practical importance that this can give extremely tight upper bounds, or even exact results. In particular, we focus on applying this approach to Hamiltonians where the Coulomb operator is diagonal, and taking the split-operator (SO) Trotterization method with second-order formulas from Ref.~\cite{kivlichan_2020}. This covers a wide range of Hamiltonians of interest for early fault-tolerant quantum computation, including: the Hubbard model and its various extensions, the Pariser–Parr–Pople model used in chemistry, the uniform electron gas (UEG), and the general electronic structure Hamiltonian in a dual-plane wave basis or on a real-space grid. Among our results, we reduce previous Trotterization costs for the UEG by an order of magnitude, and demonstrate that, for a chemistry Hamiltonian of practical interest with long-range interactions, time evolution for can be implemented with a number of Trotter steps which is sublinear in system size. We also demonstrate the use of both tensor network methods and exact quantum Monte Carlo simulation to calculate exact Trotter error norms, providing benchmark results for larger systems than can be obtained by exact calculation.

In Section~\ref{sec:upper_bound_theory} we demonstrate an upper bound on the spectral norm. We discuss the Hamiltonians and commutators of interest in Section~\ref{sec:systems}, and in Section~\ref{sec:projector_qmc} discuss a Monte Carlo method to estimate our upper bound. We then present results and discussion in Section~\ref{sec:results}.

\section{An upper bound on the spectral norm}
\label{sec:upper_bound_theory}

Consider a matrix $A$ with elements $A_{ij}$, and a matrix $\abs(A)$ with elements $|A_{ij}|$. Then,
\begin{equation}
    \lVert A \rVert \le \lVert \abs(A) \rVert,
    \label{eq:upper_bound}
\end{equation}
where $\lVert \, \cdot \, \rVert$ denotes the spectral norm, also known as the operator norm, which is equal to the largest singular value of the matrix. To prove this, note that
\begin{equation}
    \lVert A \rVert = \max_{\lVert v \rVert_2 = 1} \lVert A v \rVert_2,
\end{equation}
where $\lVert v \rVert_2$ is the vector $\ell_2$ (Euclidean) norm. Defining normalized vectors $v$ and $\abs(v)$ with $\abs(v)_i = |v_i|$,
\begin{align}
    \lVert A v \rVert_2^2 &= \sum_i \Big( \sum_j A_{ij} v_j \Big)^2, \\
    &\le \sum_i \Big( \sum_j | A_{ij} |  |v_j| \Big)^2, \\
    &= \lVert \abs(A) \, \abs(v) \rVert_2^2, \\
    &\le \lVert \abs(A) \rVert^2,
\end{align}
where the final line follows because $\abs(v)$ is a normalized vector. This shows that Eq.~(\ref{eq:upper_bound}) is true. This result has been previously discussed in the context of quantum simulation, see Ref.~\cite{childs_2010}.

Next, we state results relevant for estimating $\lVert \abs(A) \rVert$ by projector Monte Carlo simulation. Consider the case where $A$ is a square matrix. Then, $\abs(A)$ is a non-negative square matrix, and the Perron-Frobenius theorem applies \cite{meyer_2000}. Specifically, the Perron-Frobenius (PF) theorem for non-negative square matrices states that $\abs(A)$ has a real eigenvalue $\lambda_{\mathrm{PF}}$, such that
\begin{align}
    &\text{i) } \lambda_{\mathrm{PF}} \ge |\lambda_i| \;  \text{for all other eigenvalues $\lambda_i$},
    \label{eq:perrron_frobenius} \\
    &\text{ii)  The corresponding eigenvector is non-negative}.
    \label{eq:perrron_frobenius_2}
\end{align}
Therefore, $\lambda_{\mathrm{PF}}$ is the spectral radius of $\abs(A)$, defined as the maximum of the absolute values of its eigenvalues. All matrices considered in this paper will be Hermitian, and from now on we assume Hermitian matrices unless stated otherwise. In this case the spectral radius is equal to the spectral norm, so $\lambda_{\mathrm{PF}} = \lVert \abs(A) \rVert$, which can be obtained from the lowest eigenvalue of $-\abs(A)$. This result is valuable because the matrix elements of $-\abs(A)$ are all non-positive, hence this is a stoquastic and sign-problem-free matrix. The spectral norm can therefore be estimated efficiently by Monte Carlo simulation, providing a rigorous upper bound on the desired spectral norm, $\lVert A \rVert$. Note, it was shown in Ref.~\cite{spencer_2012} that this upper bound can be used to explain the origin of the sign problem in projector Monte Carlo, as discussed in Appendix~\ref{sec:sign_problem}.

For a discussion on the tightness of Eq.~(\ref{eq:upper_bound}) compared to other bounds on the spectral norm, see Ref.~\cite{childs_2010}. In the general case this bound is \emph{not} tight, particularly for dense matrices. However, our goal in this paper is to introduce a practical numerical method, and no numerical method can be accurate and efficient for all Hamiltonians. Rather, the value of any numerical method comes from its performance on a set of problems of interest. We will demonstrate that Eq.~(\ref{eq:upper_bound}) gives tight bounds on Trotter error norms for a range of important Hamiltonians.

\section{Hamiltonians and Trotter error bounds}
\label{sec:systems}

\begin{table*}[t]
\centering
{\footnotesize
\begin{tabular}{@{\extracolsep{4pt}}l@{\hskip 1mm}ccc@{\hskip 5mm}ccc}
\hline
\hline
 & \multicolumn{3}{c}{ $\lVert \, [[V,T],V] \, \rVert$ } & \multicolumn{3}{c}{ $\lVert \, [[V,T],T] \, \rVert$ } \\
\cline{2-4} \cline{5-7}
System & exact & Monte Carlo bound & $\%$ error & exact & Monte Carlo bound & $\%$ error \\
\hline
Extended Hubbard model, 1D, $N=6$ &     102.692 &      102.692 &                0.0 &              80.77 &        115.93 &                 43.5 \\
Extended Hubbard model, 1D, $N=8$ &     135.041 &      135.066 &                0.019 &            123.75 &       145.21 &                 17.3 \\
Extended Hubbard model, 1D, $N=10$ &    167.209 &      167.209 &                0.0 &              143.49 &       171.31 &                 19.4 \\
Extended Hubbard model, Hexagonal, $2\times2$ &   376.737 &      377.904 &                0.31 &             304.88 &       405.11 &                 32.9 \\
PPP model, benzene ($N=6$) &            535.593 &      535.593 &                0.0 &              775.09 &       943.45 &                 21.7 \\
PPP model, napthalene ($N=10$) &         2430.378 &     2430.387 &               0.00033 &          1780.2 &       2077.6 &                 16.7 \\
Uniform electron gas, $2\times2$ &     0.04689 &      0.04700 &                0.23 &             0.00145 &      0.00171 &                18.4 \\
Uniform electron gas, $2 \times 2 \times 2$ & $4.435 \times 10^{-4}$ & $4.437 \times 10^{-4}$ &              0.043 & $2.8 \times 10^{-4}$ & $4.2 \times 10^{-4}$ & 52.2 \\
\hline
\hline
\end{tabular}
}
\caption{Examples comparing the spectral norms of $\lVert [[V,T],V] \rVert$ and $\lVert [[V,T],T] \rVert$ (exact) to $\lVert \abs( [[V,T],V]) \rVert$ and $\lVert \abs ( [[V,T],T] ) \rVert$ (Monte Carlo bound) for a range of small systems where exact calculations can be performed. Hamiltonians are defined in Appendix~\ref{sec:hamil_definitions}. The Monte Carlo bound is extremely tight for $\lVert [[V,T],V] \rVert$ in particular, with at worst $+ 0.31 \%$ error for these examples. In some cases this upper bound is exactly tight.}
\label{tab:exact_results}
\end{table*}

We will focus on Hamiltonians $H = T + V$ where the Coulomb term, $V$, is diagonal,
\begin{equation}
    H = \sum_{ij} T_{ij} a^{\dagger}_i a_j + \sum_{i < j} V_{ij} n_i n_j.
    \label{eq:general_h}
\end{equation}
Here, $i$ and $j$ are spin orbital labels, and $n_i = a_i^{\dagger} a_i$ is a number operator for spin orbital $i$. We label the first term as the kinetic energy operator, $T$, and the second term as the potential energy operator, $V$. This class of Hamiltonians covers a large range of fermionic systems that have been the subject of numerous Trotterization studies, including the Hubbard model and its various extensions \cite{kivlichan_2018, kivlichan_2020, campbell_2022, schubert_2023, yoshioka_2024, akahoshi_2024, kan_2025, bay_smidt_2025}, and also the general electronic structure Hamiltonian in a dual plane-wave basis \cite{babbush_2018, kivlichan_2020, su_2021} or on a real-space grid \cite{low_2023, rubin_2024, berry_2025}. Hamiltonians of this form are widely studied as candidates for early fault-tolerant quantum computers with Trotterization \cite{campbell_2022, akahoshi_2024, kan_2025, bay_smidt_2025}.

Given such a Hamiltonian, we consider second-order product formulas
\begin{equation}
S_2(t) = e^{-iVt/2} e^{-i T t} e^{-iVt/2},
\label{eq:split_operator}
\end{equation}
approximating $U(t) = e^{-iHt}$, for which \cite{kivlichan_2020, childs_2021}
\begin{align}
    \lVert S_2(t) - U(t) \rVert &= \max_{|\psi\rangle} \lVert S_2(t) |\psi\rangle - U(t) | \psi\rangle \rVert_2, \label{eq:trotter_error_def} \\
    &\le W t^3,
    \label{eq:second_order_bound}
\end{align}
where we refer to $W$ as a ``Trotter error norm''. For the ordering of $S_2(t)$ above we have \cite{suzuki_1985, kivlichan_2020}
\begin{equation}
    W_{\mathrm{VTV}} = \frac{1}{12} \big\lVert [[V,T],T] \big\rVert + \frac{1}{24} \big\lVert [[V,T],V] \big\rVert,
    \label{eq:w_vtv}
\end{equation}
or for $S_2(t) = e^{-iTt/2} e^{-iVt} e^{-iTt/2}$ then
\begin{equation}
    W_{\mathrm{TVT}} = \frac{1}{24} \big\lVert [[V,T],T] \big\rVert + \frac{1}{12} \big\lVert [[V,T],V] \big\rVert.
    \label{eq:w_tvt}
\end{equation}
Eq.~(\ref{eq:second_order_bound}) is tight in that it matches the lowest-order BCH expansion term of the error operator up to a single application of the triangle inequality \cite{childs_2021}. See Appendix~\ref{sec:exact_trotter_error} for examples demonstrating the tightness of this bound. Note from Eq.~(\ref{eq:trotter_error_def}) that this error is defined as a maximum over all input states, hence assesses the \emph{worst-case} error. The average-case error \cite{zhao_2022, chen_2024}, or error evolving from a state of interest \cite{mizuta_2025}, may be much lower. We are concerned in this paper with more tightly bounding the worst-case Trotter error.

We next describe the alternative upper bounds that we will use as a point of comparison, which can be derived from the triangle inequality. First, consider the case where $T$ and $V$ both contain $\mathcal{O}(N)$ short-range terms only. As an example, consider the cuprate model with $N$ sites defined in Appendix~\ref{sec:hamil_definitions}, Eq.~(\ref{eq:cuprate_model}), where $V = U \sum_i^N n_{i\uparrow} n_{i\downarrow}$. Then,
\begin{align}
    \big\lVert [[V,T],T] \big\rVert &\le |U| \sum_i^N \big\lVert \, [[n_{i\uparrow} n_{i\downarrow}, T],T] \, \big\rVert, \label{eq:vtt_tighter} \\
    \big\lVert [[V,T],V] \big\rVert &\le |U| \sum_i^N \big\lVert \, [[n_{i\uparrow} n_{i\downarrow}, T],V] \, \big\rVert \label{eq:vtv_tighter}.
\end{align}
Both $T$ and $V$ contain $\mathcal{O}(1)$ terms acting on a given spatial orbital $i$, and therefore $[[n_{i\uparrow} n_{i\downarrow}, T],T]$ and $[[n_{i\uparrow} n_{i\downarrow}, T],V]$ act on a small $\mathcal{O}(1)$ number of sites, and the corresponding spectral norms can be calculated exactly. This approach is commonly used when considering lattice models \cite{childs_2019, childs_2021, campbell_2022, schubert_2023, bay_smidt_2025}, and we will see that it gives tight bounds in practice.

However, if $T$ and $V$ contain $\mathcal{O}(N^2)$ long-range interactions, then this approach becomes intractable without further applications of the triangle inequality. In this case, a more generally applicable approach is to write the commutator, $A$, as a linear combination of Paulis, $A = \sum_i c_i P_i$, and define the L1 norm bound
\begin{equation}
    \lVert A \rVert \le \sum_i |c_i|.
    \label{eq:l1_norm}
\end{equation}
This approach has been used, for example, in Ref.~\cite{kivlichan_2020}, which considers various Hamiltonians of the form Eq.~(\ref{eq:general_h}), including the UEG, which we compare to in this paper. This approach has the benefit that it is very general and computationally tractable, provided that the Pauli representation of the commutators can be constructed. We will show that our Monte Carlo bound also can be applied generally to Hamiltonians of the form Eq.~(\ref{eq:general_h}), but gives significantly tighter bounds.

\section{Projector Monte Carlo}
\label{sec:projector_qmc}

Our goal is to estimate the lowest eigenvalue of a sign-problem-free matrix. For this, we will use the full configuration interaction quantum Monte Carlo (FCIQMC) method \cite{booth_2009}. FCIQMC was developed for applications in quantum chemistry, and has been described in a number of papers \cite{booth_2009, spencer_2012, booth_2014, blunt_2015, blunt_2021}. Note that other QMC methods such as the Green's function Monte Carlo method \cite{becca_sorella_2017} could also be used and, particularly for sign-problem-free systems, these approaches are closely related.

Consider the task of obtaining the lowest eigenstate of a Hermitian operator, $A$, with eigenvalues $\{ \lambda_i \}$. FCIQMC achieves this by repeatedly applying an operator $P = \mathbb{1} - \Delta \tau (A - S \mathbb{1})$ to some initial state. Here, $S \mathbb{1}$ is a shift applied to the diagonal of $A$, which can be updated throughout the simulation to help control the normalization of the wave function. If $\Delta \tau < 2 / (\lambda_{\textrm{max}} - \lambda_{\textrm{min}})$ then this procedure will converge to the lowest eigenstate of $A$ upon repeated application of $P$. Defining a basis $\{ | e_i \rangle \}$ for the vector space, we write the FCIQMC wave function as $|\Psi(\tau)\rangle = \sum_i C_i(\tau) | e_i \rangle$. The corresponding update equation for the coefficients $\{ C_i(\tau) \}$ used in FCIQMC is
\begin{equation}
    C_i(\tau + \Delta \tau) = C_i(\tau) - \Delta \tau \sum_j (A_{ij} - S \delta_{ij}) C_j(\tau).
    \label{eq:update_equation}
\end{equation}
Storing the full vector of coefficients is not scalable, and so instead we store a \emph{stochastic sample} of $|\Psi(\tau)\rangle$ at each iteration, and sample terms from the update equation. Efficiently sampling terms on the right-hand-side of Eq.~(\ref{eq:update_equation}) requires that, given a basis state $|e_j\rangle$, we can efficiently sample a connected basis state $|e_i\rangle$ for which $A_{ij} \ne 0$, and can also efficiently calculate $A_{ij}$. In Appendices~\ref{sec:vtv_appendix} and \ref{sec:vtt_appendix} we describe our approach to sample from the two required commutators, $[[V,T],V]$ and $[[V,T],T]$. We show that $[[V,T],T]$ contains at most two-body excitations, and so can be sampled using existing routines optimized for electronic structure Hamiltonians. The commutator $[[V,T],V]$ is a three-body operator but has a particularly simple form which allows us to define a simple sampling algorithm, which we implemented in the HANDE-QMC code \cite{hande_paper}.

While we would like to set $A$ equal to the commutators of interest, this will have a sign problem in general. Therefore, we let $A \rightarrow - \abs(A)$ to remove the sign problem and upper bound Trotter error.

\section{Results}
\label{sec:results}

\begin{figure}[t]
\includegraphics[width=0.98\linewidth]{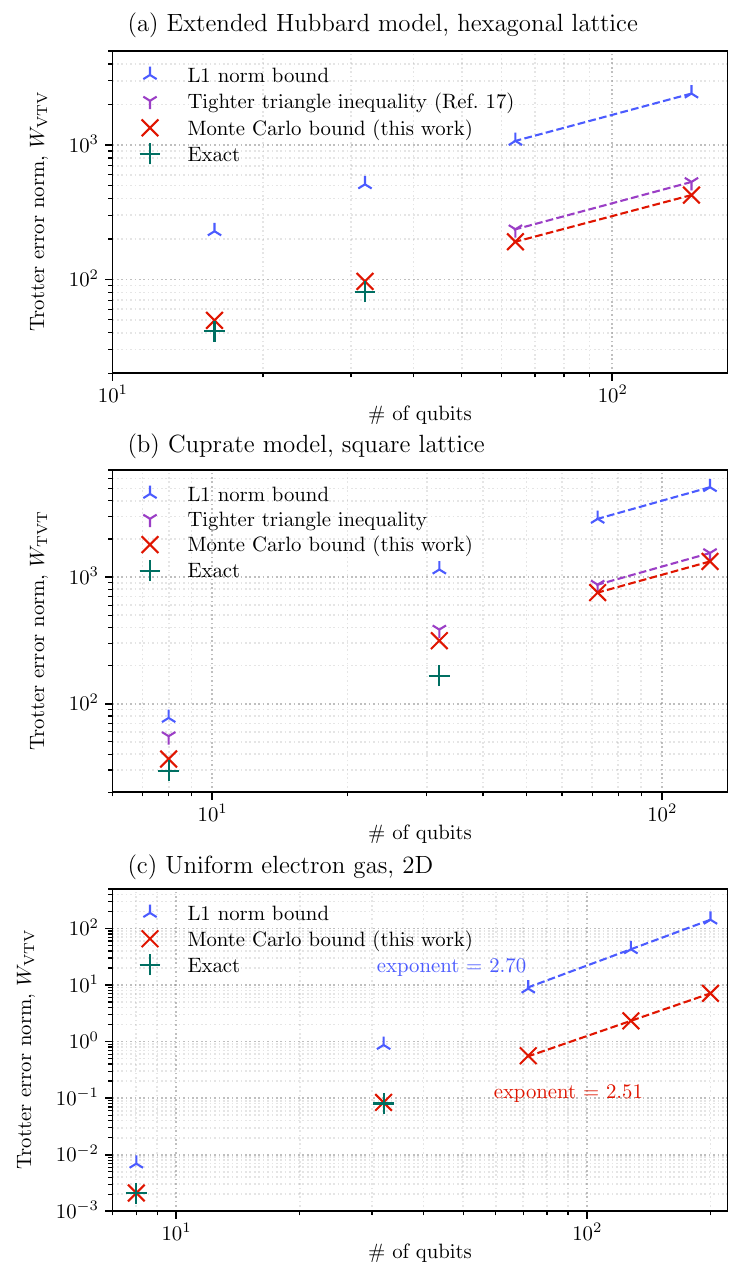}
\caption{Comparison of Trotter error norm upper bounds for two-dimensional systems. We consider: (a) the extended Hubbard model on a hexagonal lattice with $\tau=1$, $U=4$, $V=2$; (b) a cuprate model on a square lattice with $\tau=1$, $\tau'=0.3$, $\tau''=0.2$, $U=8$; and (c), the 2D uniform electron gas with $r_s = 10$. Half-filling and projected spin $s_z=0$ are taken in each case. We compare the Monte Carlo bound, Eq.~(\ref{eq:upper_bound}) to the L1 norm, Eq.~(\ref{eq:l1_norm}). We also compare to tighter triangle inequality bounds where available, using Eqs.~(\ref{eq:vtt_tighter})--(\ref{eq:vtv_tighter}) for the cuprate model and results from Ref.~\cite{bay_smidt_2025} for the extended Hubbard model. Results labeled ``Exact'' are obtained with a combination of exact diagonalization, DMRG, and exact FCIQMC (see Appendix~\ref{sec:exact_fciqmc}). Results for the UEG are in units of Ha$^3$.}
\label{fig:2d}
\end{figure}

As an initial demonstration of the bound in Eq.~(\ref{eq:upper_bound}), we take small systems where exact results can be obtained. Results are presented in Table~\ref{tab:exact_results}, considering extended Hubbard models on 1D and hexagonal lattices, the PPP model for linear acene molecules, and the UEG in 2D and 3D. The precise Hamiltonians used are defined in Appendix~\ref{sec:hamil_definitions}.

The upper bound on $\big\lVert [[V,T],V] \big\rVert$ is extremely tight, being in error by only $0.31 \%$ in the worst case considered in Table~\ref{tab:exact_results}. For the extended Hubbard model with $N=6$ sites and $N=10$ sites, and for the PPP model for benzene, the bound is exact, suggesting that $[[V,T],V]$ is sign problem free. In contrast the Monte Carlo bound on $\big\lVert [[V,T],T] \big\rVert$ is consistently looser, but still only in error by $52.2 \%$ in the worst case in Table~\ref{tab:exact_results}. Note that for second-order Trotter, the number of Trotter steps to reach a given precision goes as $\mathcal{O}(W^{1/2})$, so in this worst case the number of Trotter steps would only be overestimated by $\approx 23 \%$, and much less in practice due to the contribution from $\big\lVert [[V,T],V] \big\rVert$.

In Fig.~\ref{fig:1d} we consider 1D and quasi-1D systems. In (a) and (b) we take the PPP model applied to linear acene molecules, from benzene ($N=6$ carbon atoms) to octacene ($N=34$ carbon atoms), while in (c) and (d) we take the 1D extended Hubbard model, from $N=6$ to $N=18$ lattice sites, considering half-filling in all cases. Likely due to the linear nature of these systems, we find that DMRG is effective (though not entirely trivial to converge) in calculating the required spectral norms, which we use to provide ``exact'' benchmarks. These DMRG benchmarks were performed with the Block2 code \cite{block2}. Matching observations from Table~\ref{tab:exact_results}, the Monte Carlo bound on the spectral norm of $[[V,T],V]$ is either exact or essentially exact in each case. The $[[V,T],T]$ bounds are less accurate but still tight for practical purposes. For the PPP model the latter bound is in error by $\sim 15 \%$--$30 \%$. For the 1D extended Hubbard model, the error is $43.5 \%$ at $N=6$, but becomes tighter for larger systems, decreasing to only $9 \%$ at $N = 18$. Note that the L1 norm, Eq.~(\ref{eq:l1_norm}) leads to a significant overestimate of the Trotter error. This is particularly noticeable for the PPP model, where the scaling with $N$ is overestimated, particularly for $\big\lVert [[V,T],V] \big\rVert$.

Fig.~\ref{fig:2d} presents similar results for 2D systems, for the Trotter error norms defined in Eqs.~(\ref{eq:w_vtv}) and (\ref{eq:w_tvt}). We consider: (a) an extended Hubbard model on a hexagonal lattice (with $\tau=1$, $U=4$, $V=2$) from Ref.~\cite{bay_smidt_2025}; (b) a cuprate model on a square lattice from Ref.~\cite{kan_2025} (with $\tau=1$, $\tau'=0.3$, $\tau''=0.2$, $U=8$); and (c) the UEG (jellium) in a dual plane wave basis in 2D with Wigner-Seitz radius $r_s = 10$, as implemented in OpenFermion \cite{openfermion} (see Appendix~\ref{sec:hamil_definitions}). Note that for the cuprate model, the expected scaling of $W = \mathcal{O}(N)$ is only observed for $6 \times 6$ lattices and larger, because the model has up to third-nearest-neighbor terms. For the extended Hubbard and cuprate models, the bounds from Ref.~\cite{bay_smidt_2025} and Eqs.~(\ref{eq:vtt_tighter})--(\ref{eq:vtv_tighter}) respectively are tight in practice. However, these bounds are challenging to extend to Hamiltonians with all-to-all interactions, such as the UEG or PPP models. Fig.~\ref{fig:2d}(c) demonstrates the Monte Carlo bound on the Trotter error norm for the 2D UEG. We were able to obtain exact results for $2 \times 2$ and $4 \times 4$ grids only. We were unable to converge the latter exact result using DMRG, so instead we used an exact FCIQMC calculation (see Appendix~\ref{sec:exact_fciqmc}). For these, we see that the bound Eq.~(\ref{eq:update_equation}), calculated using sign-problem-free FCIQMC, is very tight. We perform three additional data points up to a $10 \times 10$ grid, which corresponds to a system with $200$ qubits, well beyond what could be calculated with DMRG or exact QMC methods. Note that the same UEG system was considered in Ref.~\cite{kivlichan_2020}, and we have reproduced the results therein using OpenFermion \cite{openfermion}. Notably, for the $10 \times 10$ example, we obtain $W = 7.2$ Ha$^3$, compared to $1.1 \times 10^3$ Ha$^3$ in Ref.~\cite{kivlichan_2020}, tightening this result by over $150$ times. This reduces the estimated cost of Trotterization by $\sqrt{150} \approx 12.25$ times. The discrepancy of L1 norm result between Fig.~\ref{fig:2d}(c) and Ref.~\cite{kivlichan_2020} is due to whether the L1 norm is calculated in the Jordan-Wigner or fermionic representation of the Hamiltonian.

Regarding the projector Monte Carlo method, while it can be applied to large sign-problem-free systems, there are two potential sources of error which must be checked. First, if the initial state is poor then the simulation can converge temporarily to an excited state. Second, particularly for large systems and low walker populations, a systematic bias can occur, which we discuss in Appendix~\ref{sec:population_control_bias}. Therefore, care is required to converge these simulations in practice.

Lastly, we emphasize that for second-order product formulas it suffices to perform $r= \mathcal{O}(\sqrt{W}t^{3/2}/\sqrt{\epsilon})$ Trotter steps to simulate for time $t$ with precision $\epsilon$. From the PPP results of Fig.~\ref{fig:1d}, we observe a scaling $\big\rVert [[V,T],T] \big\lVert \sim \mathcal{O}(N^{1.14})$. This suggests that
\begin{equation}
r = \mathcal{O} \Big(\frac{N^{0.57}t^{3/2}}{\epsilon^{1/2}} \Big),
\end{equation}
demonstrating sublinear scaling of $r$ with $N$ for a Hamiltonian of practical interest and with $\mathcal{O}(N^2)$ long-range interactions. For the PPP model of acenes, the two-body terms could be implemented in depth $\mathcal{O}(N)$ on a linear array of qubits using a fermionic swap network \cite{kivlichan_2018}, while the hopping terms could be implemented in $\mathcal{O}(1)$ depth using tile Trotterization \cite{bay_smidt_2025} (with $\mathcal{O}(N)$ additional Trotter error), for example.

\section{Conclusion}
\label{sec:conclusion}

We have introduced a practical numerical method to provide tighter upper bounds on Trotter error. This approach involves mapping the Trotter error norm calculation to an alternative sign-problem-free calculation, such that a rigorous upper bound on Trotter error is preserved. The sign-problem-free system can then be solved efficiently with projector Monte Carlo simulation.

Through a number of examples, considering a range of Hamiltonians of interest, we show that this mapping gives a tight upper bound on the true Trotter error norm. In particular, this approach gives significantly tighter bounds than those based on the L1 norm of the commutators, while being more generally applicable, and often tighter, than other approaches based on the triangle inequality. To demonstrate this, we calculated Trotter error norms for systems such as the Hubbard model, which have been often considered in studies of Trotter error, but also for systems with long-range all-to-all interactions, such as the PPP model and uniform electron gas. As one example, we show that Trotterization can be implemented for the PPP model of linear acene molecules with a number of Trotter steps which is sublinear in system size. These systems, particularly when treated with Trotterization, are often proposed as suitable models for early fault-tolerant devices. We therefore hope that this approach will be valuable for a range of fault-tolerant resource estimation studies.

It would be valuable to consider extending this approach to more general Hamiltonians and higher-order product formulas. The main challenge here is to efficiently sample terms from general nested commutators, and it is an interesting task to consider algorithms to accomplish this. This would bypass another problem which can occur when calculating Trotter error, which is the high memory demands in constructing such commutators directly. It would also be valuable to extend our Monte Carlo approach to sampling the average-case Trotter error \cite{zhao_2022}, rather than the worst-case error. We hope that the numerical approach presented here will allow a more accurate assessment of Trotter error in a range of interesting applications.

\section*{Data availability}

Numerical data, scripts for setting up Hamiltonians and commutators, and input and output files for DMRG and FCIQMC calculations are available on Zenodo with the DOI identifier https://doi.org/10.5281/zenodo.17341415.

\begin{acknowledgments}
We are grateful to Earl Campbell for helpful feedback on this manuscript. This work is supported by the Novo Nordisk Foundation, Grant number NNF22SA0081175, NNF Quantum Computing Programme.
\end{acknowledgments}

\bibliography{main}

\appendix

\begin{widetext}

\section{Hamiltonian definitions}
\label{sec:hamil_definitions}

In this paper we focus on Hamiltonians where the Coulomb term is diagonal, which are well-suited for Trotterization and are likely candidates for early fault-tolerant quantum computation.

The Hamiltonians considered in the main text are defined as follows. Note that here, we use $i$ and $j$ as spatial-orbital indices while explicitly writing out the spin index, $\sigma \in \{ \uparrow , \downarrow \}$.

\begin{enumerate}[i)]
    \item The extended Hubbard model with $N$ sites, including both on-site and nearest-neighbor interactions:
    \begin{equation}
        H = -\tau \sum_{\langle ij \rangle, \sigma} (a_{i \sigma}^{\dagger} a_{j \sigma} + a_{j \sigma}^{\dagger} a_{i \sigma}) + U \sum_i^N n_{i\uparrow} n_{i\downarrow} + V \sum_{\langle ij \rangle} \sum_{\sigma \sigma'} n_{i \sigma} n_{j \sigma'},
    \end{equation}
    where we set $\tau = 1$, $U = 4$, $V = 2$. We consider both one-dimensional and hexagonal lattices. This is the same system and lattices studied in Ref.~\cite{bay_smidt_2025}, to which we provide comparisons.
    
    \item An cuprate model with $N$ sites with 1st, 2nd and 3rd nearest-neighbor hopping terms:
    \begin{equation}
        H = \sum_{ij\sigma} T_{ij} a_{i \sigma}^{\dagger} a_{j \sigma} + U \sum_i^N n_{i\uparrow} n_{i\downarrow},
        \label{eq:cuprate_model}
    \end{equation}
    where $T_{ij} = -\tau$ for 1st-nearest neighbors, $T_{ij} = -\tau'$ for 2nd-nearest neighbors and $T_{ij} = -\tau''$ for 3rd-nearest neighbors. We set $\tau = 1$, $\tau' = 0.3$, $\tau'' = 0.2$, $U=8$, and consider square lattices, following the model studied in Ref.~\cite{kan_2025}.
    
    \item The Pariser–Parr–Pople (PPP) model for nanographene molecules with $N$ carbon atoms:
    \begin{equation}
        H = -\tau \sum_{\langle ij \rangle, \sigma} (a_{i \sigma}^{\dagger} a_{j \sigma} + a_{j \sigma}^{\dagger} a_{i \sigma}) + U \sum_i^N n_{i\uparrow} n_{i\downarrow} + \sum_{i<j, \sigma, \sigma'} \frac{U}{\sqrt{1 + \alpha r_{ij}^2}} n_{i\sigma} n_{j\sigma'},
    \end{equation}
    where $r_{ij}$ is the distance between atoms $i$ and $j$ in units of \AA\, and we set $U = 11.13$ eV, $t = 2.4$ eV, and $\alpha = 0.612$ \AA$^{-2}$, following one commonly-used convention \cite{Sony2009, Lambie2025}. We apply this model to linear acene molecules, where we take a fixed bond length of $1.4$ \AA. Note that while this model only has nearest-neighbor hopping terms, it has $\mathcal{O}(N^2)$ all-to-all density-density interaction terms.
    
    \item The uniform electron gas (jellium) in a dual plane wave basis~\cite{babbush_2018}:
    \begin{equation}
        H = \sum_{ij, \sigma} T_{ij} a_{i\sigma}^{\dagger} a_{j\sigma} + \sum_{(i,\sigma) > (j,\sigma')} V_{ij} n_{i\sigma} n_{j\sigma'},
    \end{equation}
    with
    \begin{equation}
        T_{ij} = 
        \sum_{\nu} \frac{k_{\nu}^2 {\cos(k_{\nu} \cdot (r_i - r_j) )}}{2N}
    \end{equation}
    and
    \begin{equation}
        V_{ij} = \sum_{\nu \ne 0} \frac{ 4\pi {\cos(k_{\nu} \cdot (r_i - r_j) )}}{ \Omega \, k_{\nu}^2 }.
    \end{equation}
    where $\Omega$ is the volume of cell, $N$ is the total number of grid points, and 
    \begin{align}
        r_i  & = \left(\frac{\Omega}{N}\right)^{1/3} (i_1, i_2, i_3)^{T}, \quad \quad i_1,i_2,i_3 \in \{0, \dots N^{1/3} -1\},\\
        k_\nu &= \frac{2\pi}{\Omega^{1/3}} (\nu_1, \nu_2, \nu_3)^{T}, \quad  \nu_1,\nu_2,\nu_3 \in \left\{-\frac{N^{1/3} }{2}, \dots ,\frac{N^{1/3}}{2} - 1\right\},
    \end{align}
    The above defines the UEG in three dimensions. Following the implementation in OpenFermion \cite{openfermion}, for the two-dimensional UEG of area $\Omega$, the matrix elements have the same analytical expression, but the grid points are sampled as
    \begin{align}
        r_i  & = \left(\frac{\Omega}{N}\right)^{1/2} (i_1, i_2)^{T}, \quad \quad i_1,i_2 \in \{0, \dots N^{1/2} -1\},\\
        k_\nu &= \frac{2\pi}{\Omega^{1/2}} (\nu_1, \nu_2)^{T}, \quad  \nu_1,\nu_2 \in \left\{-\frac{N^{1/2}}{2}, \dots ,\frac{N^{1/2}}{2} - 1\right\}.
    \end{align}
\end{enumerate}

\section{The FCIQMC algorithm}

As described in the main text, FCIQMC performs a stochastic sampling of the following update equation,
\begin{equation}
    C_i(\tau + \Delta \tau) = C_i(\tau) - \Delta \tau \sum_j (A_{ij} - S \delta_{ij}) C_j(\tau),
    \label{eq:update_equation_appendix}
\end{equation}
where $C_i(\tau)$ are coefficients of the wave function $|\Psi(\tau)\rangle = \sum_i C_i(\tau) | e_i \rangle$ in a basis $\{ | e_i \rangle \}$. In the terminology of QMC, basis states are occupied by ``walkers''. A walker is simply a non-zero coefficient on any basis state. The ensemble of walkers is updated each iteration by stochastically sampling terms on the right-hand-side of Eq.~(\ref{eq:update_equation_appendix}).

An unbiased stochastic sampling of Eq.~(\ref{eq:update_equation_appendix}) is achieved in FCIQMC by the following steps:
\begin{enumerate}
    \item Spawning: For each occupied basis state $|e_i\rangle$ with coefficient $C_i$, perform $\Nspawn$ sampling attempts. For each attempt, randomly choose a basis state $|e_j\rangle$ such that $A_{ij} \ne 0$ and $i \ne j$ with probability $P_{ij}$. Then, create a walker on state $|e_j\rangle$ with coefficient $- \Delta \tau A_{ji} C_i / (\Nspawn P_{ij})$.
    \item Death: For each occupied $|e_i \rangle$, update its coefficient by adding $- \Delta \tau A_{ii} C_i$.
    \item Merging: Sum together the existing and newly spawned walkers on each occupied basis state. Walkers of opposite sign will ``cancel out'', which is known as annihilation. Note that annihilation will not occur in sign-problem-free systems.
    \item Stochastic rounding: For all occupied basis states with a coefficient $C_i$ s.t. $|C_i| < 1$, stochastically round the coefficient to $0$ with probability $1-|C_i|$, or to $\mathrm{sign}(C_i)$ with probability $|C_i|$.
\end{enumerate}
The spawning step corresponds to the off-diagonal elements of the action of $A$ in Eq.~(\ref{eq:update_equation}), but sampling only a limited number of terms, $\Nspawn$, from the summation in an unbiased manner. A good choice for $\Nspawn$ is to round $|C_i|$ to the nearest integer. The death step corresponds to the diagonal elements of $A$, treated exactly. The merging step simply combines the newly-created walkers with the existing ones. The rounding step is important to prevent the proliferation of walkers with very low weight, which explode the memory cost; this step ensures that the minimum walker population $|C_i|$ on any occupied basis state is $1$ after each iteration.

The total walker population at a given iteration is defined by
\begin{equation}
    \Nw(\tau) = \sum_i |C_i(\tau)|,
\end{equation}
i.e. the 1-norm of the wave function. Note that, because the minimum $|C_i(\tau)|$ on any occupied $|e_i\rangle$ is $1$, this is an approximate measure of the total number of occupied states $|e_i\rangle$ that must be stored, and therefore the memory requirement of the simulation. The number of spawning attempts per iteration is also approximately equal to $\Nw$, which largely determines the runtime. Because of this, we must prevent the walker population becoming too large. If the shift $S$ is not set sufficiently low then the walker population will grow exponentially with the iteration number. However, if it is set too low then the walker population will decay to zero. To avoid this, the shift is updated after each iteration using
\begin{equation}
    S(\tau+\Delta\tau) = S(\tau) - \frac{\xi}{\Delta \tau} \mathrm{ln} \Bigg( \frac{\Nw(\tau + \Delta \tau)}{\Nw(\tau)} \Bigg),
    \label{eq:shift_update}
\end{equation}
in order to counteract changes in the walker population. Here, $\xi$ is a parameter called the ``shift damping'', determining how aggressively $S$ is updated. Assuming that the ground state is sufficiently converged upon, the growth rate of $\Nw(\tau)$ is determined by the lowest eigenvalue of $A$. Therefore, by varying the shift to counteract changes in $\Nw(\tau)$, the shift $S(\tau)$ can be used as an estimator of the lowest eigenvalue. Note that varying the shift leads to a bias in the simulation, known as the population control bias. This bias is usually small but can become significant for large enough systems. We discuss this bias in Appendix~\ref{sec:population_control_bias}.

\begin{figure*}[t]
\includegraphics[width=1.0\linewidth]{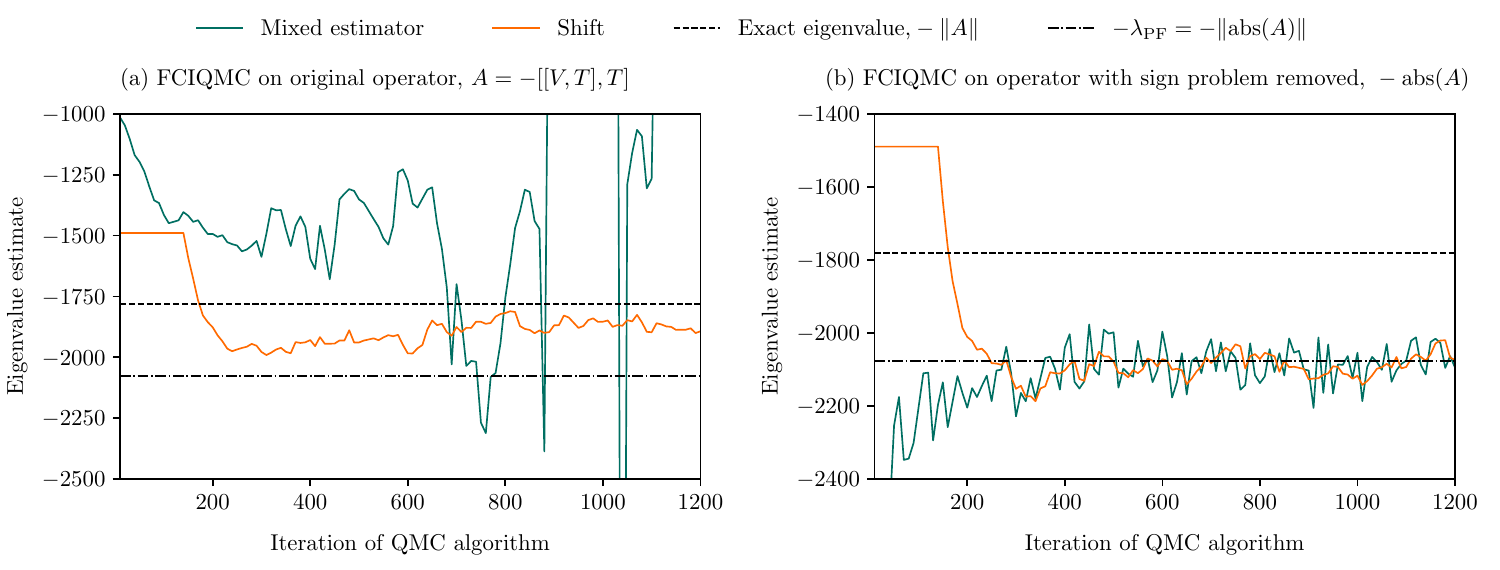}
\caption{FCIQMC simulations performed on the commutator $A = -[[V,T],T]$ for a napthalene molecule described by the PPP model. The dashed line is minus the spectral norm, and the dashed-dotted line is minus the spectral norm of the sign-problem-free commutator, denoted $-\lambda_{\mathrm{PF}}$. (a) FCIQMC performed on the exact commutator. This operator has a sign problem, so the shift estimator converges to a value below $- \lVert A \rVert$, and the signal from the mixed estimator becomes swamped in noise beyond iteration $\sim 250$. (b) FCIQMC performed on the sign-problem-free operator, $-\abs(A)$. Here the propagation is stable and both estimators sample $-\lVert \abs(A) \rVert$, where $\lVert \abs(A) \rVert \ge \lVert A \rVert$ is guaranteed. In both (a) and (b) the shift is allowed to vary once the walker population (not shown) reaches $1000$. Note that for this example we could sample the exact solution in (a) with a slightly larger walker; we use a low walker population here for demonstrative purposes.}
\label{fig:fciqmc_example}
\end{figure*}

Another estimator of the lowest eigenvalue is the mixed estimator,
\begin{align}
    \lambda_{\mathrm{mixed}}(\tau) &= \frac{\langle \Psi_{\mathrm{trial}} | A | \Psi(\tau) \rangle }{\langle \Psi_{\mathrm{trial}} | \Psi(\tau) \rangle}, \\
    &= \frac{ \sum_{ij} \psi_i A_{ij} C_j(\tau) }{\sum_i \psi_i C_i(\tau)},
    \label{eq:mixed_estimator}
\end{align}
where $|\Psi_{\mathrm{trial}}\rangle = \sum_i \psi_i |e_i\rangle$ is a trial solution to the lowest eigenvector of $A$. For the commutators studied in this paper, coming up with an accurate trial wave function would be challenging, since commutators do not directly correspond to standard Hamiltonians. However, since we remove the sign problem, and using the property in Eq.~(\ref{eq:perrron_frobenius_2}), we can simply take $\psi_i = 1$, i.e., an equally weighted positive superposition of all states:
\begin{equation}
    \lambda_{\mathrm{mixed}}(\tau) = \frac{ \sum_{ij} A_{ij} C_j(\tau) }{\Nw(\tau)},
    \label{eq:final_mixed_estimator}
\end{equation}
where we have used that $\sum_i C_i = \sum_i |C_i| = \Nw$ for the sign-problem-free systems in this paper. Lastly, note that the numerator of Eq.~(\ref{eq:final_mixed_estimator}) can be sampled directly from the spawned walkers, since the spawned walkers perform an unbiased sampling of the right-hand-side of Eq.~(\ref{eq:update_equation_appendix}), which contains $\sum_j A_{ij} C_j(\tau)$. Therefore, we can obtain a sampling of the mixed estimator defined in Eq.~(\ref{eq:final_mixed_estimator}) for no additional cost, which can be used for all commutators using the sign-problem-free approach that we have defined. Note that sampling the numerator of Eq.~(\ref{eq:final_mixed_estimator}) in this manner may result in large statistical errors. For chemistry calculations, where we seek chemical accuracy, this could be a limiting factor. However, for estimating Trotter error norms we typically do not need such small error bars, and we find this approach to work well.

In Fig.~\ref{fig:fciqmc_example} we present an example, applying FCIQMC to (a) the operator $A = -[[V,T],T]$ and (b) the operator $- \abs( \,[[V,T],T] \,)$ for a napthalene molecule within the PPP model. Note that the eigenvalue of $[[V,T],T]$ largest in magnitude for this system is the highest eigenvalue; since FCIQMC obtains the lowest eigenvalue, we simulate $A = -[[V,T],T]$ rather than $A = +[[V,T],T]$. The mixed energy estimator (green) in (a) becomes unstable beyond the initial iterations, while the shift estimator (orange) converges to a value below the true eigenvalue. In (b) both estimators are stable and converge to $-\lambda_{\mathrm{PF}}$, where $\lambda_{\mathrm{PF}} = \lVert \, \abs( \,[[V,T],T] \,) \, \rVert$. For this example we find $\big\lVert \, \abs\big( [[V,T],T] \big) \, \big\rVert = 1.167 \, \big \lVert \,[[V,T],T] \, \big\rVert$,
giving a tight bound in error by only $16.7\%$.

\section{The sign problem in FCIQMC}
\label{sec:sign_problem}

The sign problem in FCIQMC was discussed in detail in Ref.~\cite{spencer_2012}. This is directly related to Eq.~(\ref{eq:upper_bound}), and so we briefly recap this connection here. This sign problem can be explained by considering the FCIQMC algorithm \emph{without} annihilation. In this case, the positive and negative walkers are separate. Define the positive walker coefficients by $C_i^+$, and the negative coefficients by $C_i^-$. Likewise, define the matrix $Q_{ij} = \langle e_i | S \mathbb{1} - A | e_j \rangle$, and define a matrix of positive elements by $Q_{ij}^+$ and absolute values of negative elements by $Q_{ij}^-$, i.e., such that $Q = Q^+ - Q^-$. Then, \emph{in the case where positive and negative walkers are not allowed to annihilate}, the following equations can be derived (in the limit of small $\Delta\tau$) \cite{spencer_2012},
\begin{align}
    \frac{d (C_i^+ + C_i^-)}{d\tau} = \sum_j \big( Q_{ij}^+ + Q_{ij}^-\big) \big(C_j^+ + C_j^- \big), \\
    \frac{d (C_i^+ - C_i^-)}{d\tau} = \sum_j \big( Q_{ij}^+ - Q_{ij}^-\big) \big(C_j^+ - C_j^- \big).
\end{align}
Therefore, the desired signal $C^+ - C^-$ converges to the highest eigenvector of $Q$ (which is the highest eigenvector of $-A$ or the lowest eigenvector of $A$), as desired. However, the signal $C^+ + C^-$ will converge to the highest eigenvector of $\abs(A)$, whose eigenvalue is $\lambda_{\mathrm{PF}} = \lVert A \rVert$. From the results of Section~\ref{sec:upper_bound_theory} we have $\lambda_{\mathrm{PF}} \ge |\lambda_{\mathrm{min}}|$. This is a problem because $C^+ + C^-$ determines the walker population. In practice, the walker population cannot be allowed to grow exponentially indefinitely, but must eventually be stabilized to control memory requirements. The shift, $S$, must be set as $S \approx -\lambda_{\mathrm{PF}}$ to achieve this. Since $- \lambda_{\mathrm{PF}} \le \lambda_{\mathrm{min}}$,  we have $S - \lambda_{\mathrm{min}} \le 0$, and the signal from the lowest eigenstate will necessarily decay to zero and be lost in statistical noise (except for sign-problem-free systems where $-\lambda_{\mathrm{PF}} = \lambda_{\mathrm{min}}$).

In FCIQMC we allow annihilation between positive and negative walkers, which suppresses the growth rate of the walker population, counteracting the above effect. For small enough systems, or large enough walker populations, stable sampling can be achieved even for systems which are not sign problem free. In Appendix~\ref{sec:exact_fciqmc} we demonstrate such FCIQMC simulations to exactly calculate Trotter error norms.

\section{Algorithm to sample $[[V,T],V]$}
\label{sec:vtv_appendix}

\subsection{Derivation of commutator expression}

In this Appendix we derive an expression for $[[V,T],V]$, where $T$ and $V$ are one-body and diagonal two-body operators, respectively, considering Hamiltonians of the form in Eq.~(\ref{eq:general_h}).

In our derivation of the nested commutator expressions, we use the following commutator rules for fermionic creation ($a^\dagger$), annihilation ($a$) and number ($n = a^\dagger a$) operators, which hold for our case of interest where spin-orbital labels $i$, $j$ and $k$ obey $i \neq j$ and $k \neq i,j$:
\begin{eqnarray}
    & & [n_i n_j, a^\dagger_i a_j] = 0, \label{eq:comm1}\\
    & & [n_i n_k, a^\dagger_i a_j] = n_k a^\dagger_i a_j, \label{eq:comm2}\\
    & &[n_i n_k, a^\dagger_j a_i]=  -n_k a^\dagger_j a_i. \label{eq:comm3}
\end{eqnarray}
In the case where $i = j$, $a_i^\dagger a_j = a_i^\dagger a_i  = n_i$ for which $[V, n_i] = 0$.

We consider $[[V, T],V]$, which can be expressed as
\begin{equation}
    [[V, T],V] = \sum_{i < j} T_{i j} [[V, a^\dagger_i a_j + a^\dagger_j a_i],V], \label{eq:commChC_initial}
\end{equation}
where the diagonal of $T$ does not contribute as it commutes with $V$. We evaluate two parts of the commutator separately using (\ref{eq:comm1}), (\ref{eq:comm2}) and (\ref{eq:comm3}),
\begin{eqnarray}
    & & [V, a^\dagger_i a_j] =  \sum_{k \neq j, i}^{2N}  \Big ( V_{ik} - V_{jk}\Big)n_k  a^\dagger_i a_j \label{eq:Vaidagaj}, \\
    & & [V, a^\dagger_j a_i] =  \sum_{k \neq j, i}^{2N} \Big ( V_{jk} - V_{ik} \Big) n_k  a^\dagger_j a_i \label{eq:Vajdagai},
\end{eqnarray}
and $[V, a^\dagger_i a_j + a^\dagger_j a_i]$ can be obtained by adding (\ref{eq:Vaidagaj}) and (\ref{eq:Vajdagai}). We nest the resulting expression with $V$ and use that $n_k$ commutes with all terms in $V$ 
\begin{eqnarray}
    & & [[V, a^\dagger_i a_j + a^\dagger_j a_i ],V] = \sum_{k \neq j, i}^{2N} \Bigg( \Big( V_{ik} - V_{jk} \Big) n_k [a^\dagger_i a_j, V] +  \Big( V_{jk} - V_{ik} \Big)  n_k [a^\dagger_j a_i, V] \Bigg).
\end{eqnarray}
Using that $[A,B] = -[B,A]$ and Eqs.~(\ref{eq:Vaidagaj}) and (\ref{eq:Vajdagai}), we obtain
\begin{eqnarray}
    [[V,a^\dagger_i a_j + a^\dagger_j a_i ],V] &=& -\sum_{k \neq j, i}^{2N} \sum_{q \neq j, i}^{2N} \Bigg( \Big( V_{ik} - V_{jk} \Big)\Big( V_{iq}-V_{jq} \Big) n_k n_q  a^\dagger_i a_j +  \Big( V_{jk} - V_{ik} \Big)\Big( V_{jq}-V_{iq} \Big) n_k n_q  a^\dagger_j a_i \Bigg) \nonumber \\
    &=& -\sum_{k \neq j, i}^{2N} \sum_{q \neq j, i}^{2N} \Big( V_{ik} - V_{jk} \Big)\Big( V_{iq}-V_{jq} \Big) n_k n_q  (a^\dagger_i a_j + a^\dagger_j a_i).
\end{eqnarray}
Using (\ref{eq:commChC_initial}), the final commutator may be written as
\begin{equation}
    [[V, T ],V] = -\sum_{i < j} \Bigg( T_{ij} \sum_{k \neq j, i}^{2N} \sum_{q \neq j, i}^{2N} \Big( V_{ik} - V_{jk} \Big)\Big( V_{iq}-V_{jq} \Big) n_k n_q  (a^\dagger_i a_j + a^\dagger_j a_i) \Bigg).
\end{equation}
We can further rearrange this to obtain
\begin{align}
    [[V, T ],V] &= -\sum_{i < j} \Bigg( T_{ij} (a^\dagger_i a_j + a^\dagger_j a_i) \sum_{k \neq j, i}^{2N} \Big( V_{ik} - V_{jk} \Big) n_k \sum_{q \neq j, i}^{2N} \Big( V_{iq}-V_{jq} \Big) n_q \Bigg), \\
    &= -\sum_{i < j} T_{ij} (a^\dagger_i a_j + a^\dagger_j a_i) \Bigg( \sum_{k \neq j, i}^{2N} \Big( V_{ik} - V_{jk} \Big) n_k \Bigg)^2.
    \label{eq:vtv_expression}
\end{align}

\subsection{Matrix elements and sampling of of $[[V,T],V]$}
\label{sec:vtv_sampling}

To perform FCIQMC on the commutator $[[V,T],V]$, we need to be able to efficiently calculate matrix elements between and also sample excitations from one basis state to another. We perform FCIQMC in the fermionic representation using a basis of Slater determinants $\{ |D_i\rangle \}$, hence we need to calculate matrix elements of the form
\begin{equation}
    \langle D_i | \, [[V,T],V] \, | D_j \rangle.
\end{equation}

The expression for $[[V,T],V]$ was given in Eq.~(\ref{eq:vtv_expression}). As can be seen, this is a three-body operator, however, it \emph{only} couples determinants which are a single excitation apart. The diagonal elements are also equal to $0$. Therefore, we only need to find an expression for single excitations, and the Monte Carlo code only needs to generate single-body excitations.

Denote a Slater determinant by $|D\rangle$. Then, let the notation $|D_i^{\,j}\rangle$ denote a second determinant which differs from $|D\rangle$ by exciting a single electron from a previously occupied orbital $i$ into a previously unoccupied orbital $j$.

Then, starting from Eq.~(\ref{eq:vtv_expression}), we can simplify the matrix element $\langle D_i^{\,j} | \, [[V,T],V] \, | D \rangle$ as follows,
\begin{align}
    \langle D_i^{\,j} | \, \Big[ [V,T],V \Big] \, | D \rangle &= - T_{ij} \, \langle D_i^{\,j} | a^{\dagger}_j a_i | D\rangle \, \langle D | \left( \sum_{k \neq j, i}^{2N} \Big( V_{ik} - V_{jk} \Big) n_k \right)^2 | D \rangle, \\
    &= - T_{ij} \, \langle D_i^{\,j} | a^{\dagger}_j a_i | D\rangle \, \left( \sum_{k \in |D\rangle, k \ne i} \Big( V_{ik} - V_{jk} \Big) \right)^2,
\end{align}
where the summation in the second line is over all spin orbitals occupied in $|D\rangle$ except for $i$ (note that $j$ is unoccupied in $|D\rangle$ by definition). When performing Monte Carlo on the sign-problem-free operator, where we take the absolute values of matrix elements, we obtain
\begin{equation}
    \Big| \langle D_i^{\,j} | \, \big[ [V,T],V \big] \, | D \rangle \Big| = |T_{ij}| \, \left( \sum_{k \in |D\rangle, k \ne i} \Big( V_{ik} - V_{jk} \Big) \right)^2.
\end{equation}
This matrix element can be calculated in $\mathcal{O}(N)$ time, where $N$ is the number of electrons.

Since only single excitations lead to non-zero matrix elements, we can perform excitation generation in FCIQMC using existing routines optimized for molecular Hamiltonians, where both single and double excitations must be generated. After first randomly choosing an occupied spin orbital $i$ to excite from, we next randomly choose an occupied spin orbital $j$ of the same spin to excite to. We choose an orbital of the same spin because all Hamiltonians considered in this paper have spin symmetry, where $T_{ij} = 0$ if $i$ and $j$ have differing spin.

Note that this could be optimized further by first randomly choosing an occupied orbital $i$, and then randomly choosing an orbital $j$ such that $T_{ij} \ne 0$. This slightly slows down the simulation, because calculating the probability $P_{\mathrm{gen}}$ of having selected the excitation (which is necessary to reweight the amplitude of the spawned walker) becomes more expensive. However it would reduce statistical noise by removing generation of null excitations where $T_{ij} = 0$. Nonetheless, we find the above excitation generation sufficient for the systems studied in this paper.

\section{Derivation of commutator $[[V, T],T]$}
\label{sec:vtt_appendix}

We next show that $[[V, T],T]$ contains only up to two-body operators. This commutator can be expressed as
\begin{equation}
    [[V, T],T] = \sum_{i<j} V_{ij} [[n_in_j,T],T]. \label{eq:commChh_initial}
\end{equation}
We begin by evaluating $[n_i n_j, T]$ using (\ref{eq:comm1}), (\ref{eq:comm2}) and (\ref{eq:comm3}),
\begin{equation}
    [n_i n_j, T] = \sum_{k\neq i,j}^{2N} \Bigg( T_{ik} n_j (a^\dagger_i a_k - a^\dagger_k a_i) + T_{jk} n_i (a^\dagger_j a_k - a^\dagger_k a_j) \Bigg).
\end{equation}
We proceed by nesting the above commutator expression with $T$ and expanding using the commutator identity $[AB,C] = A[B,C] + [A,C]B$,
\begin{align}
    [[n_i n_j, T],T] &= \sum_{k\neq i,j}^{2N} \Big( T_{ik} [n_j (a^\dagger_i a_k - a^\dagger_k a_i),T] + T_{jk} [n_i (a^\dagger_j a_k - a^\dagger_k a_j),T] \Big), \nonumber \\
    &= \sum_{k\neq i,j}^{2N} \Big ( T_{ik} n_j [a^\dagger_i a_k - a^\dagger_k a_i,T] + T_{ik} [n_j, T](a^\dagger_i a_k - a^\dagger_k a_i) + T_{jk} n_i [a^\dagger_j a_k - a^\dagger_k a_j,T] + T_{jk}[n_i, T](a^\dagger_j a_k - a^\dagger_k a_j) \Big), \nonumber \\
    &= \sum_{k\neq i,j}^{2N} \Big (T_{ik} n_j [a^\dagger_i a_k - a^\dagger_k a_i,T] + T_{ik} \sum_{q\neq j}^{2N} T_{jq} (a^\dagger_j a_q - a^\dagger_q a_j) (a^\dagger_i a_k - a^\dagger_k a_i) \nonumber \\ &\;\;\;\;\;\;\;\;\;\;\; + T_{jk} n_i [a^\dagger_j a_k - a^\dagger_k a_j,T] + T_{jk} \sum_{p\neq i}^{2N} T_{ip} (a^\dagger_i a_p - a^\dagger_p a_i)(a^\dagger_j a_k - a^\dagger_k a_j) \Big).
\end{align}
Note, if $j \neq q$, then $[n_j, a^\dagger_j a_q]=a^\dagger_j a_q$ and $[n_j, a^\dagger_q a_j]=-a^\dagger_q a_j$. This expression contains up to two-body terms since the commutator of two one-body operators is another one-body operator. Using (\ref{eq:commChh_initial}), $[[V, T],T]$ can be written as
\begin{align}
    [[V, T],T] &= \sum_{i<j} V_{ij}\sum_{k\neq i,j}^{2N} \Big (T_{ik} n_j [a^\dagger_i a_k - a^\dagger_k a_i,T] + T_{ik} \sum_{q\neq j}^{2N} T_{jq} (a^\dagger_j a_q - a^\dagger_q a_j) (a^\dagger_i a_k - a^\dagger_k a_i) \nonumber \\ &\;\;\;\;\;\;\;\;\;\;\;\;\;\;\;\;\;\;\;\;\;\; + T_{jk} n_i [a^\dagger_j a_k - a^\dagger_k a_j,T] + T_{jk} \sum_{p\neq i}^{2N} T_{ip} (a^\dagger_i a_p - a^\dagger_p a_i)(a^\dagger_j a_k - a^\dagger_k a_j) \Big).
\end{align}

We could use this expression to construct the commutator $[[V,T],T]$ directly and then construct an algorithm to sample this operator, similarly to the approach discussed in Section~\ref{sec:vtv_appendix}. However, since this operator contains up to two-body terms, the total number of terms in a given $[[V,T],T]$ is lower than compared to $[[V,T],V]$. For the systems considered in this paper we can easily construct the $[[V,T],T]$ commutators directly, which we perform using OpenFermion \cite{openfermion}. Excitations and corresponding matrix elements of $[[V,T],T]$ can then be sampled in FCIQMC using existing excitation generators designed for electronic structure Hamiltonians \cite{booth_2014}, which also contain at most two-body terms. Specifically, we use the excitation generators implemented for electronic structure Hamiltonians as implemented in the HANDE-QMC software package \cite{hande_paper}.

\section{Comparison to exact worst-case Trotter error}
\label{sec:exact_trotter_error}

In this Appendix we present results comparing exact worst-case Trotter error to the commutator bounds and our Monte Carlo estimates of the commutator bounds for three representative small systems, where the exact Trotter error can be easily calculated.

As discussed in the main text, we consider a Hamiltonian $H = V + T$ and a product formula $S_2(t) = e^{-iVt/2} e^{-iTt} e^{-iVt/2}$ approximating $U(t) = e^{-iHt}$. We define the worst-case Trotter error as
\begin{equation}
    \lVert S_2(t) - U(t) \rVert = \max_{|\psi\rangle} \lVert S_2(t) |\psi\rangle - U(t) | \psi\rangle \rVert_2.
    \label{eq:exact_trotter_error_def}
\end{equation}
We then define upper bounds on the exact Trotter error as \cite{childs_2021}
\begin{equation}
    W t^3 = \frac{t^3}{12} \big\lVert [[V,T],T] \big\rVert + \frac{t^3}{24} \big\lVert [[V,T],V] \big\rVert
    \label{eq:w2_app}
\end{equation}
and the bound from this work which can be calculated from Monte Carlo simulation,
\begin{equation}
    W_{\mathrm{MC}} t^3 = \frac{t^3}{12} \big\lVert \abs( [[V,T],T] ) \big\rVert + \frac{t^3}{24} \big\lVert \abs( [[V,T],V] ) \big\rVert,
    \label{eq:w2_mc}
\end{equation}
which obey
\begin{equation}
    \lVert S_2(t) - U(t) \rVert \le W t^3 \le W_{\mathrm{MC}} t^3.
    \label{eq:three_bounds_app}
\end{equation}

Results are presented in Fig.~\ref{fig:exact_trotter_error}. We consider: the PPP model applied to benzene; the extended Hubbard model on a 1D lattice with $N=6$ sites; and the 2D uniform electron gas with $r_s=10$ and a $2 \times 2$ grid. These are all small examples where the true worst-case error can easily be obtained. We find that both Eq.~(\ref{eq:w2_app}) and Eq.~(\ref{eq:w2_mc}) are extremely tight upper bounds for the UEG example. For the PPP model and extended Hubbard model, these bounds are slightly more loose, but still tight overall, always within a factor of less than $2$ times. We emphasize that the number of Trotter steps to perform time evolution to a given precision goes as the square root of this error for second-order Trotterization, and so the total cost is only slightly overestimated. Also, the Monte Carlo Trotter error norm $W_{\mathrm{MC}}$ will become a tighter upper bound on $W$ for larger systems, because the $\lVert [[V,T],V] \rVert$ dominates for larger systems, for which our Monte Carlo bound is particularly tight. As noted in the main text, we emphasize that the \emph{average-case} Trotter error may be much lower in practice \cite{zhao_2022, chen_2024}.

\begin{figure*}[t]
\includegraphics[width=1.0\linewidth]{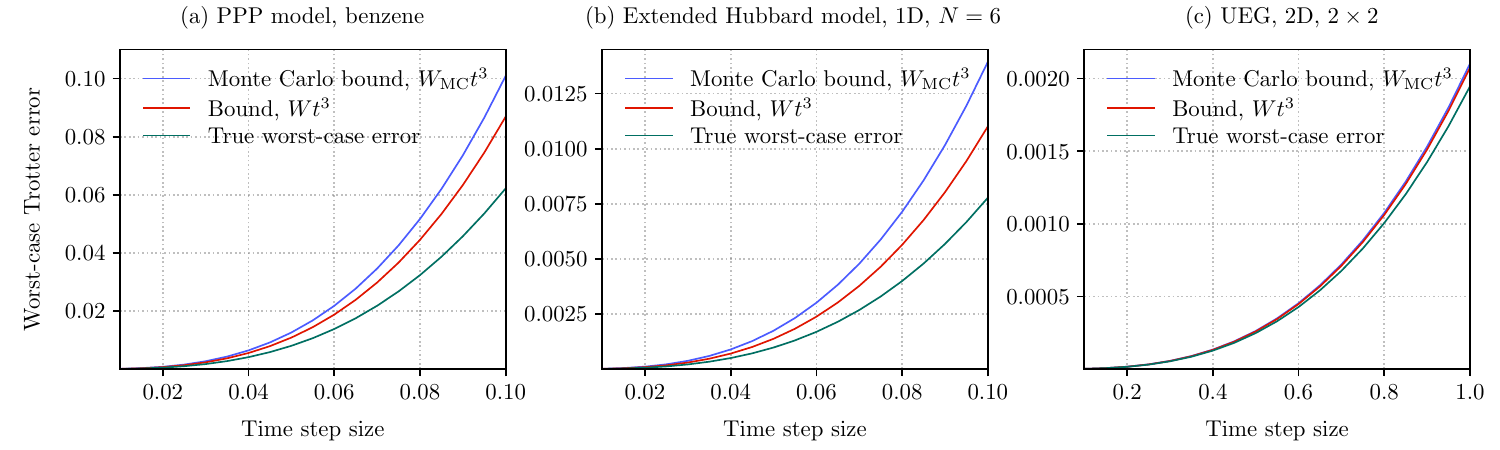}
\caption{Results comparing the true worst-case Trotter error (green) of Eq.~(\ref{eq:exact_trotter_error_def}), to the commutator bound (red) of Eq.~(\ref{eq:w2_app}) and the bound which can be calculated efficiently by Monte Carlo (blue) of Eq.~(\ref{eq:w2_mc}). We consider three systems from the main text, for small examples where exact Trotter error can be calculated: (a) the PPP model for a benzene molecule; (b) the extended Hubbard model on a 1D lattice with $N=6$ sites; (c) the uniform electron gas in 2D, with $r_s=10$ and a $2 \times 2$ grid.}
\label{fig:exact_trotter_error}
\end{figure*}

\section{Further results for two-dimensional systems}
\label{sec:2d_systems}

\begin{figure}[t]
\includegraphics[width=1.0\linewidth]{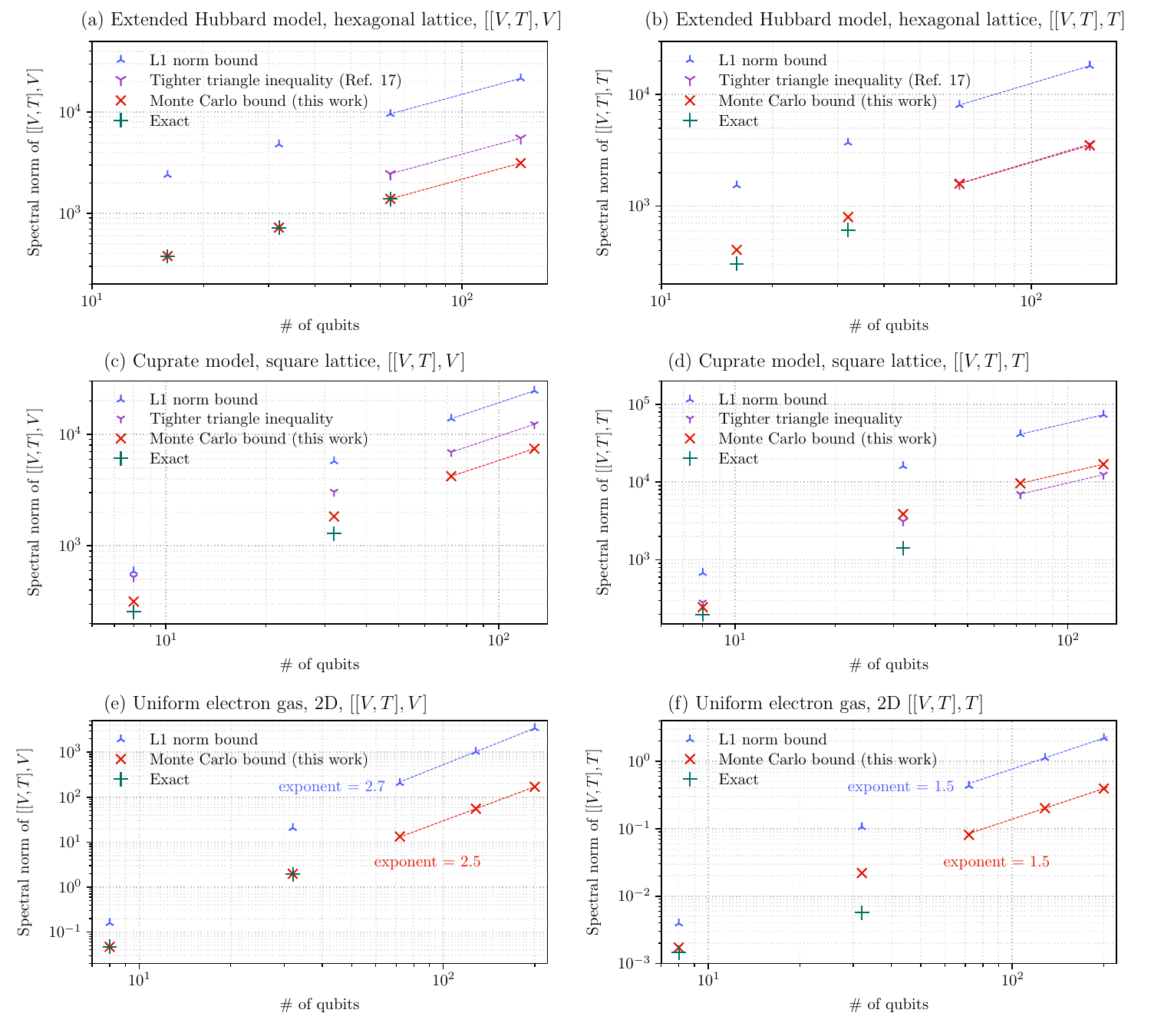}
\caption{Comparison of spectral norm estimates (restricted to half-filling and projected spin $s_z=0$) for commutators $[[V,T],V]$ and $[[V,T],T]$, for (a--b) the extended Hubbard model on a hexagonal lattice with $t=1$, $U=4$, $V=2$; (c--d) a cuprate model on a square lattice with $t=1$, $t'=0.3$, $t''=0.2$, $U=8$; and (e--f), the uniform electron gas with $r_s = 10$. These results correspond to those from Fig.~\ref{fig:2d} in the main text. Results for the UEG are in units of Ha$^3$. Dashed lines are included to guide the eye for larger systems, where finite-size effects are smaller.}
\label{fig:2d_all}
\end{figure}

In this Appendix we provide further data to accompany Fig.~\ref{fig:2d} in the main text. In Fig.~\ref{fig:2d}, we presented the Trotter error norms, $W_{\mathrm{VTV}}$ and $W_{\mathrm{TVT}}$, defined as
\begin{equation}
    W_{\mathrm{VTV}} = \frac{1}{12} \big\lVert [[V,T],T] \big\rVert + \frac{1}{24} \big\lVert [[V,T],V] \big\rVert,
    \label{eq:w_vtv_app}
\end{equation}
and
\begin{equation}
    W_{\mathrm{TVT}} = \frac{1}{24} \big\lVert [[V,T],T] \big\rVert + \frac{1}{12} \big\lVert [[V,T],V] \big\rVert,
    \label{eq:w_tvt_app}
\end{equation}
corresponding to $S(t) = e^{-iVt/2} e^{-iTt} e^{-iVt/2}$ and $S(t) = e^{-iTt/2} e^{-iVt} e^{-iTt/2}$, respectively. In our calculations we also restrict the spectral norm to half-filling and the projected spin $s_z=0$ sector, although this is not necessary in general.

Here, we additionally plot results for $\big\lVert [[V,T],T] \big\rVert$ and $\big\lVert [[V,T],V] \big\rVert$ to assess the tightness of the Monte Carlo-based bound for the individual commutators. Results are presented in Fig.~\ref{fig:2d_all}, for the same systems and data as in Fig.~\ref{fig:2d}.  From this additional data, we highlight a few points. First, as noted in the main text, the Monte Carlo-based bound on $\big\rVert [[V,T],V] \big\lVert$ is particularly tight. The bound on $\big\rVert [[V,T],T] \big\lVert$ is somewhat less tight, but still accurate for practical purposes. The exception is for the uniform electron gas in Fig.~\ref{fig:2d_all}(f), where the Monte Carlo bound for the latter commutator and a $4 \times 4$ grid is higher than the benchmark results by a factor of $\sim 3.9$ times. However, because $\big\rVert [[V,T],V] \big\lVert$ is orders of magnitude larger than $\big\rVert [[V,T],V] \big\lVert$ for the UEG system, the final estimate of $W_{\mathrm{VTV}}$ is extremely tight. However, this does hint at possible examples where the Monte Carlo bound may be less accurate. A second point to highlight is in Fig.~\ref{fig:2d_all}(d), where the ``tighter triangle inequality'' bound from Eq.~(\ref{eq:vtt_tighter}) is slightly tighter than the Monte Carlo bound. However, since these are both rigorous upper bounds, we can always take the lower of the two. We emphasize that this triangle inequality bound is challenging to extend to non-lattice Hamiltonians, such as the UEG example, where the Monte Carlo bound from this work is still simple to calculate.

Although this paper has focused on a numerical approach to calculate prefactors for Trotter error, we can also compare to asymptotic results. For Hamiltonians of the same form that we consider, Ref.~\cite{low_2023} derived the following asymptotic bound for a general $p$'th-order product formula,
\begin{equation}
    \lVert S_p(t) - e^{-iHt} \rVert_{\eta} = \mathcal{O} \Big( ( \vertiii{T}_1 + \vertiii{V}_{1,[\eta]} )^{p-1} \vertiii{T}_1 \vertiii{V}_{1,[\eta]} \eta t^{p+1} \Big),
    \label{eq:asymptotic_upper_bound}
\end{equation}
where $\lVert \, \cdot \, \rVert_\eta$ is the spectral norm restricted to the $\eta$-particle subspace, $\vertiii{T}_1$ is the induced $1$-norm and $\vertiii{V}_{1,[\eta]}$ is the restricted induced $1$-norm,
\begin{equation}
    \vertiii{T}_1 = \max_i \sum_j |T_{ij}|,
\end{equation}
\begin{equation}
    \vertiii{V}_{1,[\eta]} = \max_i \max_{j_1 < \ldots < j_{\eta}} ( |V_{i, j_1}| + \ldots |V_{i, j_{\eta}}| ).
\end{equation}
This result has been used to perform asymptotic complexity analysis of higher-order product formulas \cite{low_2023, berry_2025}, but also resource estimation with prefactors by estimating the constant of proportionality for small systems \cite{rubin_2024}. We compare this asymptotic result to our results for the two-dimensional uniform electron gas. By numerically evaluating $\vertiii{T}_1$ and $\vertiii{V}_{1,[\eta]}$ up to $40 \times 40$ grids (at which point the scaling is well converged) at half-filling ($\eta = N$), we find $\vertiii{T}_1 = \mathcal{O}(1)$ and $\vertiii{V}_{1,[\eta]} = \mathcal{O}(N)$. For $p=2$ product formulas this suggests that the Trotter error norm obeys $W = \mathcal{O}(N^3)$. This compares to the numerical fit from our Monte Carlo-based upper bounds, which suggests $W = \mathcal{O}(N^{2.5})$. Hence, we believe that the asymptotic upper bound of Eq.~(\ref{eq:asymptotic_upper_bound}) still overestimates the scaling slightly for this Hamiltonian. Nonetheless, Eq.~(\ref{eq:asymptotic_upper_bound}) holds for any product formula and so is clearly a valuable result.

\section{Using FCIQMC to obtain exact commutator bounds}
\label{sec:exact_fciqmc}

In the main text we focused on using FCIQMC to obtain upper bounds on spectral norms of commutators, as required to provide upper bounds for Trotter error. To do this, we used the bound from Eq.~(\ref{eq:upper_bound}) to transform a given matrix to a sign-problem-free matrix, where FCIQMC can be performed efficiently.

However, FCIQMC can also be used to obtain \emph{exact} solutions for systems with sign problems, albeit with a memory and time cost that is exponential in system size. For systems that are not too large, this can be manageable. Indeed, this was the original motivation for FCIQMC \cite{booth_2009}, which has been applied to sign-problem-free systems more rarely.

In this appendix we provide examples of using FCIQMC to obtain exact commutator spectral norms, for two systems where we were unable to converge DMRG satisfactorily.

FCIQMC is able to perform stable sampling of ground-state solutions in the presence of a sign problem due to walker annihilation. Annihilation is when walkers of opposite sign cancel out and are removed from the simulation. As explained in detail in Ref.~\cite{spencer_2012}, in the absence of annihilation the walker growth rate is determined by the largest eigenvalue of $\abs(A)$, when performing FCIQMC on an operator $A$. This is necessarily larger than the highest eigenvalue of $-A$, which determines the growth rate of the desired state. This results in the sign problem. However, walker annihilation reduces the growth rate of the walker population without affecting the growth rate of the desired solution. At high enough annihilation rates (obtained at high walker populations), this can allow stable sampling, even in the presence of a sign problem.

\begin{figure*}[t]
\includegraphics[width=1.0\linewidth]{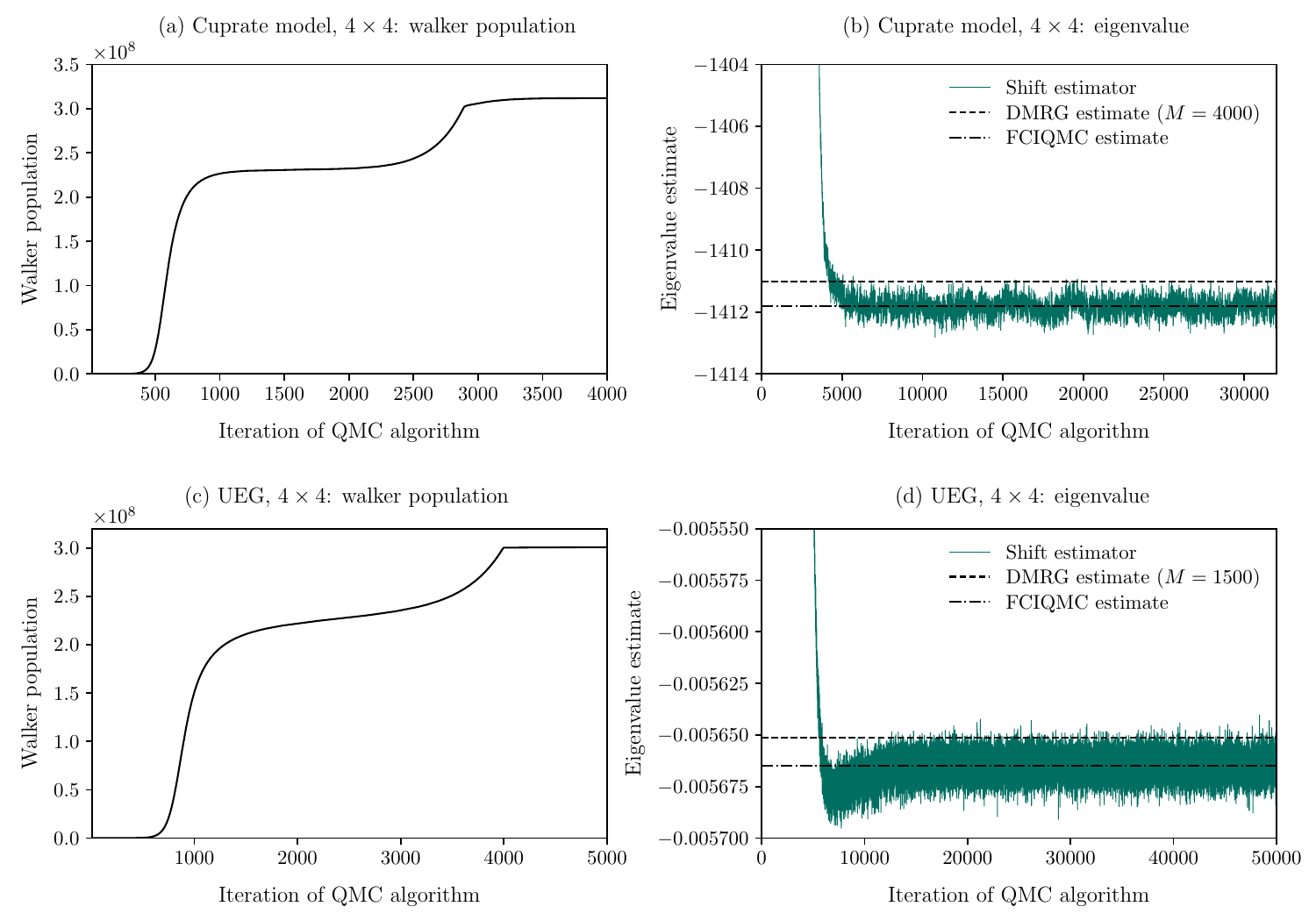}
\caption{Two example FCIQMC simulations to calculate $\lVert [[V,T],T] \rVert$, \emph{without} first removing the sign problem. Subplots (a) and (b) come from a single FCIQMC simulation, performed on $A = -[[V,T],T]$ for the cuprate model on a $4 \times 4$ square lattice. Subplots (c) and (d) come from a single FCIQMC simulation, performed on $A = -[[V,T],T]$ for the uniform electron gas (UEG) in 2D at $r_s=10$ and with a $4 \times 4$ grid. In (a) and (c) the walker population is seen to enter plateau period at around $2 \times 10^8$ walkers in both simulations. After leaving this plateau, the exact solution can be stably sampled. At walker population $N_w = 3 \times 10^8$ we being varying the shift, $S$, to stabilize the walker population. Subplots (b) and (d) show the value that the shift subsequently converges to. The mean shift estimate (after convergence) is labeled ``FCIQMC estimate''. Also shown are DMRG estimates using large bond dimensions, $M$, which were not fully converged with respect to $M$. Eigenvalue estimates for the UEG are in units of Ha$^3$.}
\label{fig:exact_fciqmc}
\end{figure*}

In Fig.~\ref{fig:exact_fciqmc}, we show two example FCIQMC simulations demonstrating exact calculation (within small statistical errors) of the lowest eigenvalue of $-\lVert [[V,T],T] \rVert$ for two systems: the cuprate model on a $4 \times 4$ square lattice, and the two-dimensional UEG using a $4 \times 4$ grid. Both of these systems would be represented by $32$ qubits under a Jordan-Wigner mapping, for example. For clarity, we emphasize that subplots (a) and (b) come form a single FCIQMC calculation, and subplots (c) and (d) also come from a single simulation.

Subplots (a) and (c) show the walker population dynamics in the two simulations. At the start of the simulation we fix the shift, $S$, so that the walker population is allowed to grow freely. Initially this population grows exponentially with imaginary time (iteration number). However, as the population grows, the walker annihilation rate also grows, until eventually walker spawning and annihilation rates cancel out, leading to a ``plateau'' period. During this plateau the desired ground-state signal emerges, and so upon leaving the plateau we can begin to vary the shift, $S$, to control the walker population. Subplots (b) and (d) present the respective values of the shift, which we use as an estimator for the lowest eigenvalue the commutator. A detailed explanation of walker dynamics and the sign problem in FCIQMC is presented in Ref.~\cite{spencer_2012}.

In these plots we also show DMRG estimates of the same eigenvalue, using the largest bond dimension, $M$, which we could perform using the Block2 code \cite{block2}. These simulations were not fully converged with respect to $M$. In contrast, using the same node with the same memory constraints, we were able to sample the exact solutions using FCIQMC for these systems. However, we believe it is likely that DMRG could be converged, either by performing a calculation with more memory, or a more advanced setup of the simulation. Also, while the DMRG estimates here will not give a strict upper bound on the Trotter error, the error is by less than $1 \%$ in both cases. For most practical purposes of estimating Trotter error, such a small error will likely not be a problem, particularly as Trotter error norms are expected to be somewhat loose. However, for benchmarking purposes it is useful to be able to provide converged estimates.

\section{Population control bias in FCIQMC}
\label{sec:population_control_bias}

This paper uses a projector Monte Carlo method to find the ground-state solution of sign-problem-free systems. While projector QMC is considered to be efficient for this task, there are a small number of places where biases can appear in such simulations. Perhaps the most challenging example of this is population control bias. This is a well-known phenomena in all projector QMC methods, including diffusion Monte Carlo \cite{umrigar_1993}, Green's function Monte Carlo \cite{runge_1992}, and FCIQMC \cite{vigor_2015, ghanem_2021, brand_2022}. In brief, this bias appears because the shift parameter, $S$, is updated to counteract changes in the walker population, as defined in Eq.~(\ref{eq:shift_update}). As a result, the weight of particular configurations in the final energy estimator will be either suppressed or amplified (relative to a simulation where $S$ is held constant). In particular, if the walkers enter a low-energy configuration then the shift will be made more negative, decreasing the walker population. If the walkers enter a high-energy configuration then the shift will be made more positive, increasing the walker population. As a result, the weight of lower-energy configurations will be decreased, while the weight of higher-energy configurations will be increased. This essentially reweights contributions in the energy estimator, Eq.~(\ref{eq:mixed_estimator}), such that energies will become systematically overestimated.

This bias is usually negligible in FCIQMC simulations, and other projector QMC methods. However, when considering very large Hilbert spaces, this bias can become significant such that it must be carefully removed. Experience shows that this bias mostly becomes significant in sign-problem-free systems \cite{runge_1992, ghanem_2021}, in part because very large Hilbert spaces can be sampled in this case.

We have found that, for the largest systems studied in this paper, the population control bias in the Trotter error commutators can become significant, and larger than previously observed in FCIQMC, to the best of our knowledge. This seems to be particularly true for the $[[V,T],V]$ commutator.

There are perhaps two main approaches to remove the population control bias. The first is by directly reweighting the energy estimator to ``undo'' the effect of having varied the shift \cite{runge_1992, umrigar_1993}. This is generally a good and widely used approach, however, it can significantly increase the statistical error in the energy estimator. To remove the bias to a satisfactory degree, it may be necessary to include reweighting factors from many thousands or tens of thousands of iterations prior. This might require very significant averaging (i.e., millions of iterations) to properly converge the resulting statistical errors. A second approach is to obtain an estimate of the energy at several different walker populations and to perform an extrapolation. The population control bias must decrease with increasing walker populations, which allows such extrapolations to be performed.

In our simulations, we found it challenging to satisfactorily converge population control bias using the reweighting procedure. As such, we have instead used extrapolations to approximately remove this bias, when necessary. For most calculations performed, the bias was negligible such that this was not necessary. However, particularly for simulations on the $[[V,T],V]$ commutator for systems over $\sim 100$ qubits, we find that the bias can become significant.

\begin{figure*}[t]
\includegraphics[width=1.0\linewidth]{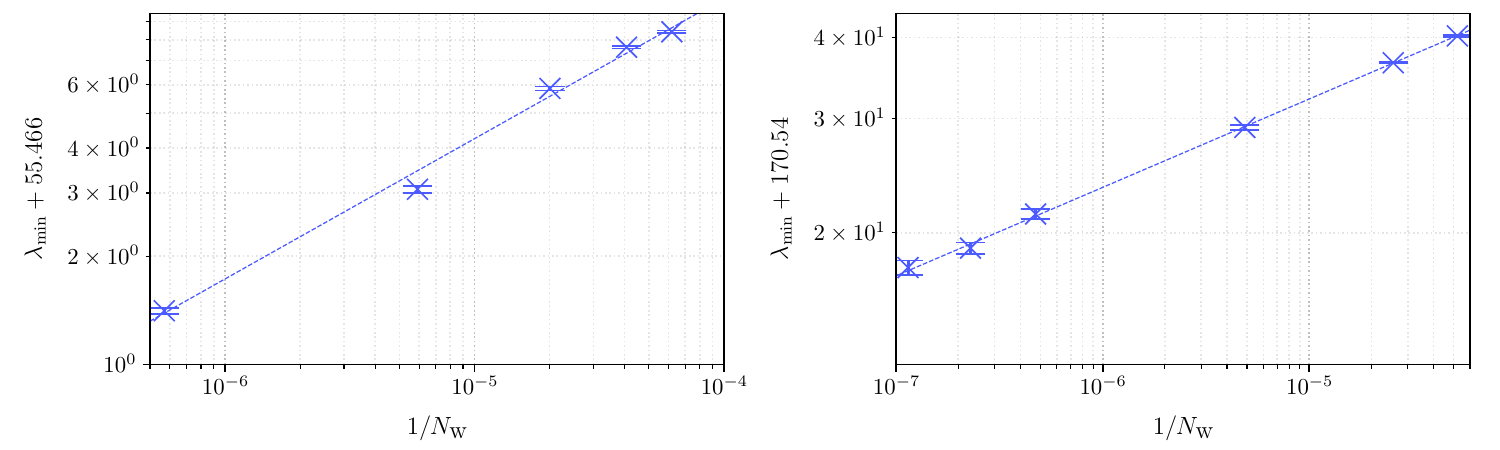}
\caption{Estimates of the lowest eigenvalue of the $-\abs( \, [[V,T],V] \,)$ operator, for the UEG on an $8 \times 8$ grid (left) and $10 \times 10$ (right). On the X-axis is the reciprocal of the average walker population, $\Nw$, in each simulation. On the Y-axis we plot the estimate of the lowest eigenvalue, shifted by our extrapolated estimate of the true eigenvalue, in units of Ha$^3$. This extrapolation is done by fitting to the function $y = a + bx^c$, and taking $a$ as the estimate at $N_\mathrm{w}=0$ (the infinite walker population limit). For the $10 \times 10$ grid in particular, the extrapolation is $\sim 10 \%$ of the spectral norm estimate, so is important to perform.}
\label{fig:pop_control_bias}
\end{figure*}

In Fig.~\ref{fig:pop_control_bias}, we present this extrapolation for the two most challenging examples considered in this paper, the two-dimensional UEG with $8 \times 8$ (left) and $10 \times 10$ (right) grids. The corresponding systems, after performing a qubit mapping, would have $128$ and $200$ qubits, respectively, which are well beyond direct calculation by exact methods. We perform FCIQMC simulations at a range of different $\Nw$ values, up to $1.75 \times 10^6$ for the $8 \times 8$ case, and up to $8.78 \times 10^6$ for the $10 \times 10$ case. We obtain energies from the mixed estimator in Eq.~(\ref{eq:mixed_estimator}) (which is known to have lower population control bias than the shift estimator). We then fit these values to a function of the form $y = a + b x^c$ using a SciPy routine, which also gives an estimate of the error on the parameters. We then take $\lambda_{\mathrm{min}} = a$ as our estimate in the infinite walker population limit. This fitting form is ultimately a hypothesis, however, the population control bias has been shown to follow a power law decay through extensive experience in the projector QMC literature. For the $8 \times 8$ case we obtain $\lambda_{\mathrm{min}} = -55.5(7)$ Ha$^3$, and for the $10 \times 10$ case we obtain $\lambda_{\mathrm{min}} = -171(5)$ Ha$^3$. The estimate of the spectral norm is then $\lVert \, \abs( [[V,T],V] ) \, \rVert = -\lambda_{\mathrm{min}}$.

For the $10 \times 10$ example in particular, the size of the extrapolation is large. The population control bias at $\Nw = 8.78 \times 10^6$ remains around $18(5)$ Ha$^3$, which is around $10 \%$ of the estimated unbiased value. However, there are a few reasons to believe that this issue can be significantly improved. Firstly, the walker populations (and computational resources) used here are relatively low, compared to state-of-the-art QMC calculations. The $10 \times 10$ UEG results were performed with only $8.78 \times 10^6$ walkers on $40$ MPI processes. By comparison, FCIQMC can be performed with billions of walkers and parallelizes efficiently to at least tens of thousands of MPI processes. So, if desired, much larger systems could be studied using more computational resources. Secondly, there are ways to reduce population control bias. Most obviously, this can be done by varying the shift less aggressively, i.e., by reducing $\xi$ in Eq.~(\ref{eq:shift_update}). As noted at the end of Appendix~\ref{sec:vtv_sampling}, better excitation generators can also be used for $[[V,T],V]$, which we expect to reduce this bias. In the longer term, there may be other QMC methods better suited for sign-problem-free simulation than FCIQMC, such as continuous-time QMC algorithms. Lastly, we note that, while the size of the extrapolation in Fig.~\ref{fig:pop_control_bias} for the $10 \times 10$ example would be concerning for quantum chemistry applications (where we seek chemical accuracy), for estimating Trotter error, a small final uncertainty is not necessarily a problem in practice, unless the goal is to set benchmark numbers, and these estimates of Trotter error are still expected to be somewhat loose.

\end{widetext}

\end{document}